*Monthly Notices of the Royal Astronomical Society*

# A Monte Carlo based simulation of the galactic chemical evolution of the Milky Way galaxy.

Sandeep Sahijpal and Tejpreet Kaur, Dept. of Physics, Panjab University, Chandigarh, India 160014 (sandeep@pu.ac.in)

## Abstract

The formation and chemical evolution of the Milky Way Galaxy is numerically simulated by developing a Monte Carlo approach to predict the elemental abundance gradients and other galactic features using the revised solar abundance. The galaxy is accreted gradually by using either a two-infall or a three-infall accretion scenario. The galaxy is chemically enriched by the nucleosynthetic contributions from an evolving ensemble of generations of stars. We analyse the role of star formation efficiency. The influence of the radial gas inflow as well as radial gas mixing on the evolution of galaxy is also studied. The SN Ia delay time distribution (DTD) is incorporated by synthesizing SN Ia populations using random numbers based on a distribution function. The elemental abundance evolutionary trends corroborate fractional contributions of ~0.1 from prompt (< 100 Myr.) SN Ia population. The models predict steep gradients in the inner regions and less steep gradients in the outer regions which agrees with the observations. The gradients indicate an average radial gas mixing velocity of $\leq 1$ km s$^{-1}$. The models with radial gas inflows reproduce the observed inversion in the elemental abundance gradients around 2 billion years. The three-infall accretion scenario performs better than the two-infall accretion model in terms of explaining the elemental abundance distributions of the galactic halo, thick and thin discs. The accuracy of all the models has been monitored as a cumulative error of < 0.15 M$_\odot$ in the mass balance calculations during the entire evolution of the galaxy.

## 1.      Introduction

The origin of the Milky Way galaxy presumably commenced within the initial one billion years (Gyr) subsequent to the Big-Bang origin of the Universe around 13.7 Gyr ago. The formation of the galaxy initiated with the merging of the initial diffuse neutral hydrogen gas clouds. This was followed by the gravitational merging of the colliding initial protogalaxies, thereby, leading to the gradual accretional evolution of the galaxies in a hierarchical manner. The successive generations of stars formed and evolved inside the galaxy led to the gradual abundance evolution of the elemental (and isotopic) inventories of the galaxy. The galactic chemical evolution (GCE) models deal with understanding the origin of the distribution and gradual evolution of the elemental (and isotopic) abundances across the galaxy over the galactic timescale of ~13.7 Gyr (Matteucci and François 1989; Rana 1991; Chiappini, Matteucci & Gratton 1997; Goswami & Prantzos 2000; Chiappini, Matteucci & Romano 2001; Pagel 1997; Chang et al. 1999; Alibés, Labay & Canal 2001; Matteucci 2003, 2014; François et al. 2004; Kobayashi et al. 2006; Kobayashi and Nomoto 2009; Matteucci et al. 2009; Kobayashi and Nakasato 2011; Kobayashi, Karakas & Umeda 2011; Minchev, Chiappini & Martig 2013, 2014; Sahijpal & Gupta 2013; Sahijpal 2013, 2014; Micali, Matteucci & Romano 2013; Mott, Spitoni & Matteucci 2013; Sahijpal 2014; Minchev et al. 2015; Spitoni et al. 2015). The astronomical observed trends in the elemental abundance distribution across the galaxy serve as a major constraint on the GCE models in order to understand the accretional growth rate and the dynamical evolution of the galaxy with its three main components, i.e., the galactic halo, the thick disc and the thin disc (Marochnik & Suchkov 1996; Matteucci 2003; Sparke & Gallagher 2007; Chiappini et al. 2015). The GCE model provides a detailed evolutionary account of the star formation rate (*SFR*), the average prevalent initial mass function (*IMF*) for star formation, the supernovae (SN Ia, Ib/c, II) rates and the stellar nucleosynthetic contributions of stars with distinct mass and metallicity.

The GCE models can be broadly categorized into two types on the basis of the adopted numerical approach. The majority of the conventional GCE models involve solving the integro-differential equations dealing with the isotopic abundance evolution of the elements over the galactic timescale (see e.g., Matteucci & Greggio 1986; Matteucci and François 1989; Pagel 1997; Matteucci 2003, 2014). An assumed galactic accretion rate, the SFR and the IMF are employed along with the prescribed stellar nucleosynthetic contributions from stars of different mass and metallicity in order to solve the equations. At the local solar neighborhood these simulations aim to achieve the solar metallicity ($Z_\odot$) value along with the elemental (isotopic) abundance of the solar system at the time of the formation of the solar system around 4.56 Gyr ago. However, the GCE models across the galactic spatial extent deal with understanding the origin of the observed abundance gradients of different elements and their temporal evolution (see e.g., Chiappini et al. 1997, 2001; Matteucci 2003; Cescutti et al. 2007; Colavitti et al. 2009, Micali et al. 2013, Mott et al. 2013; Spitoni et al. 2015). Further, in addition to understand the intricacies related with the formation and the evolution of the galactic halo, the thick and thin discs, the GCE models are essential to understand the dynamical processes related with stellar migration and radial gas mixing across the galaxy.

Apart from the conventional GCE models several N-body numerical models, referred as the chemodynamical models, incorporate the dynamics along with the galactic chemical evolution in a self-consistent manner. These models use smoothed particle hydrodynamics (SPH) technique to incorporate the galactic dynamics. In the majority of these models each simulation particle represents $10^4$-$10^5$ $M_\odot$ instead of a single star. The supernova feedback and chemical enrichment processes are incorporated in the hydrodynamical code to simulate the kinematic and chemical properties of the galactic components (Fux 1997, 1999, 2001a, b; Gibson et al. 2003; Kobayashi 2004, 2005; Kobayashi & Nakasato 2011). The recent chemodynamical models (Minchev et al. 2012a, b, 2013, 2014) address the important issues related with the influence of stellar migration on the evolution of the galaxy. The astronomical observations of the age-metallicity relation indicate that stars drift away from their birth place. Stellar migration significantly influences the chemical and kinematical properties of the galactic disc (Sellwood & Binney 2002; Haywood 2008; Roškar et al. 2009; Spitoni et al. 2015). The radial migration results from the change in the angular momentum and epicycle motion of stars because of their interaction with spiral arms (Schönrich & Binney 2009). Stellar migration caused by the resonance in the galactic bar and spiral arms has been also hypothesized to explain the dispersion in the age-metallicity relation (Minchev & Famaey 2010).

In addition to stellar migration, the radial flow of intragalactic gas also plays an important role in influencing the observed elemental abundance trends. The radial gas flow has been proposed as one of the reasons behind the observed elemental abundance gradients along the disc (Tinsley & Larson 1978). However, certain models indicate that gas flows alone cannot explain the entire observed elemental gradients (Götz & Köppen 1992). The accretion of the infalling galactic matter brings changes in the angular momentum of the disc which leads to the radial gas flows. The flow is further enhanced by the gravitational interaction between the spiral arms and the molecular gas clouds, the collision among the interstellar gas clouds and the viscosity differences among the gas layers (Tinsley & Larson 1978; Lacey & Fall 1985; Sellwood & Binney 2002; Schönrich & Binney 2009; Spitoni & Matteucci 2011). The velocity of the radial gas flows has been proposed to be in the range of 0.1-1 km s$^{-1}$ (Portinari & Chiosi 2000; Spitoni et al. 2015). The observations of H I emission and absorption features suggest an upper limit of few km s$^{-1}$ for the gas inflows along the disc (Lacey & Fall 1985). Further, the observational evidences indicate inversion in the elemental gradients at high red shift that correspond to ~2 billion years after the Big Bang (Cresci et al. 2010). The inverted abundance gradients have been explained by adopting specific radial gas inflow in the galaxy (Mott et al. 2013).

One of the major objectives of the GCE models is to understand the origin and evolution of stars in the galactic halo, the thick and thin discs. The stellar population observations of these components suggest diverse evolutionary history. The general adopted approach for the galaxy formation involves the two-infall accretion model (Chiappini et al. 1997). The galactic halo was formed in the first episode. The second episode resulted in the formation of galactic disc gradually from the intergalactic gas. The thin disc formed comparatively at a longer timescale which is considered to be a function of the galactocentric distance. This resulted in an inside-out accretion scenario according to which the inner regions of the disc were formed earlier and the outer regions were formed later on (Matteucci & François 1989; Chiappini et al. 2001). There is an alternative suggestion proposed by Colavitti, Matteucci & Murante (2008) and Colavitti et al. (2009) based on the cosmological infall law which does not predict the inside-out criteria for the disc formation. There have been alternative hypotheses for the accretion of matter in the galaxy, e.g., the models with Gaussian infall accretion rate (Chang et al. 1999), and the three stage infall accretion model (Micali et al. 2013). The models with an exponentially declining accretional rates are considered to be most successful. In general, the underlying mechanism for the formation of the galactic thick and thin discs remains debatable.

The spectroscopic observations of the alpha-nuclide elements assist in decoding the formation history of different components of the galaxy. Observations predict a short timescale for the formation of the central Milky Way bulge. An identical trend found in the local thick disc favors a similar chemical evolution (Elmegreen 1999; Meléndez et al. 2008; Babusiaux et al. 2010). Observations also indicate distinct chemical and kinematic properties of the stellar populations belonging to thick and thin discs (Gilmore & Reid 1983; Bensby & Feltzing 2010). The stars belonging to thick disc are relatively old and alpha-nuclide rich with higher velocity dispersion (Schonrich & Binney 2009). Even though, the formation scenario of the thick disc is still debatable, some N-body simulations suggest that the thick disc evolved from the thickening of the galactic thin disc due to internal heating that happened because of radial migration, turbulence and gravitational instabilities. The thick disc evolution at high redshift supports this scenario (Bournaud, Elmegreen & Martig 2009; Bovy, Rix & Hogg 2012).

The galactic radial gas flow is considered as one of the important factors in determining the GCE. It influences the chemical evolution as the elements synthesized inside the stars can be carried out to distant regions by galactic wind, thereby, altering the abundance distribution of the disc. In general, the galactic outflow is considered to be proportional to SFR (Matteucci 2003, 2014; Finlator & Davé 2008; Dalla & Schaye 2008; Recchi et al. 2008).

One of the major emphasis of developing the GCE model is to explain the origin of the elemental (isotopic) composition of the solar system. Recently, the solar metallicity ($Z_\odot$) has been revised from a value of ~ 0.020 to 0.0143 (Anders & Grevesse 1989; Asplund el al. 2009). The elemental abundance of the sun represents the standard elemental abundance distribution for the stellar evolution and nucleosynthetic theories. It also serves as a standard reference for comparison among the observed stellar elemental abundance distribution. The recent significant revision in the solar metallicity has a substantial influence on the deduced GCE models. The N-body numerical simulations for the solar neighborhood developed on the basis of the evolution of successive generations of stars over the galactic timescale indicate a distinct evolution on account of the reduction in the solar metallicity (Sahijpal & Gupta 2013; Sahijpal 2013, 2014).

N-body Monte Carlo based numerical simulations have been developed in the present work to understand the evolution of the Milky Way galaxy across the radial extent from 2-18 kpc in terms of an evolving ensemble of stars that are formed and evolved within the galaxy according to the prevailing surface mass density and star formation rates. The main objective of the present work is to understand the nature of the accretional growth of the galaxy in terms of its three components, viz, the galactic halo, thick and thin discs, and deduce the SFRs, the supernovae (SN Ia, Ib/c, II) rates, the elemental (isotopic) abundance evolution across the galaxy over the galactic timescale. We deduce the elemental abundance gradients across the galaxy at distinct timesteps during its evolution. The numerical simulations are performed with the revised solar metallicity (Asplund et al. 2009) for a specific set of simulation parameters in order to understand the influence of the proposed two-infall and three-infall accretion scenarios, the radial gas inflow, the radial gas mixing and the SFR on the GCE. The predicted results from the theoretical models are compared with the spectroscopic observations of Planetary Nebulae (PNe), Cepheids, Red Giants, Open Clusters (OCs), OB stars, etc. (see e.g., Chiappini et al. 2001; Minchev et al. 2014; Spitoni et al. 2015; Anders et al. 2017). These observational elemental abundance trends give insight into the galactic formation history and impose constraints on the simulation parameters.

Compared to the conventional approach based upon solving the integro-differential equations (e.g., Pagel 1997; Matteucci 2003, 2014) that forms the basis for the majority of the GCE models proposed earlier, except for the chemodynamical models, the present developed procedure offers a substantially simplified numerical approach based on an evolving ensemble of stars that are formed and evolved within the galaxy over the galactic timescale. This approach needs no prerequisite consideration regarding the intricacies involved in the analytical approach in the conventional GCE models. The numerical procedure can be developed directly in a manner based on Monte Carlo simulations to understand the chemical evolution of the galaxy. Apart from presenting a detailed in-depth analysis of the chemical evolution of the Milky Way galaxy, we demonstrate that the numerical approach has the potential to incorporate all the features of the conventional GCE models and to an extent the dynamical aspects of the chemodynamical models. Even though we can numerically incorporate the influence of stellar migration on GCE, we have restricted our discussion to radial gas mixing and radial gas inflow in the present work to confine the length of the manuscript. We present a detailed study of the temporal evolution of several elemental abundance gradients. The preliminary results of this work were presented in an international conference (Sahijpal & Kaur 2018). The role of stellar migration on GCE will be discussed in our future work. Nonetheless, we will provide a qualitative understanding regarding the influence of stellar migration based on our simulations. Finally, it should be mentioned that we have not employed hydrodynamical model in our code. This is one of the shortcomings of the present work. However, we can incorporate the dynamical issues related with gas mixing and inflow in a numerical manner.

The procedure adopted in the numerical simulations of the galactic chemical evolution is discussed in the section 2 along with the assumptions in the adopted SFR, the galaxy accretion scenario, the IMF and the stellar nucleosynthetic details. The simulation results based on a selective set of critical parameters and evolutionary scenarios are presented in section 3. The major findings of this work along with the comparison with the previous GCE models are discussed in section 4. Finally, the work is summarized in section 5.

## 2. Numerical Simulations

The major objective of any GCE model irrespective of the involved analytical or numerical techniques (see e.g., Chiappini et al. 1997; Kobayashi & Nakasato 2011; Minchev et al. 2013, 2014; Brusadin, Matteucci & Romano 2013; Sahijpal & Gupta 2013; Matteucci 2014) is to numerically simulate the evolution of the galaxy to understand its formation and the gradual elemental (isotopic) abundance evolution. The conventional approach is based upon solving the integro-differential equations involving various isotopic species (Chiappini et al. 1997; Pagel 1997; Matteucci 2014) and the chemodynamical models integrate the dynamical evolution with the chemical evolution of the galaxy (Kobayashi & Nakasato 2011; Minchev et al. 2013, 2014; Brusadin et al. 2013). The current developed approach is based on the N-body Monte Carlo simulations that deal with the evolution of successive generations of stars in the Milky Way galaxy. We have developed the simulations for the evolution of the galaxy across its radial extent from 2-18 kpc. The galaxy is radially divided into eight annular rings of width 2 kpc each. We initiate the simulations with a prescribed formulation for the formation of the galaxy. The galaxy is accreted gradually from the intergalactic gas following an exponentially decreasing infall rate. An evolving ensemble of billions of simulation stars with mass in the range of 0.1-100 $M_\odot$ are formed across the galaxy over the galactic timescale from the accreted matter that has been recycled by the steady stellar nucleosynthetic contributions from the evolved simulation stars. This is achieved by defining a stellar number distribution function, $G_i(t, m)$. This function represents the total number of stars corresponding to a particular mass, $m$, formed within the i[th] annular galactic ring according to the initial mass function (IMF) and the prevailing star formation rate, *SFR*, at a specific time, *t*. The stars of different mass and metallicity defined by this function evolve over distinct timescales. Subsequent to their evolution, the stellar nucleosynthetic yields are instantaneously homogenized with the interstellar gas over the specific annular ring. Thus, the metallicity ($Z$) of the interstellar medium evolves distinctly for different annular rings over time. The simulations are performed in a manner so that at the time of the formation of the solar system around 4.56 Gyr ago the solar neighborhood acquires the revised solar metallicity ($Z_\odot$) value of ~0.0143 (Asplund el al. 2009) and [Fe/H] ~0. This epoch is defined to be around 9 billion years after the initiation of the formation of the galaxy.

We adopted either the conventional two or three stage infall model for the accretion of the galactic halo-thick disc and thin disc (Chiappini et al. 1997; Brusadin et al. 2013; Micali, et al. 2013; Matteucci 2014). The eight annular rings of the simulated galaxy are numbered from 1 to 8, starting from the inner most ring. The solar neighborhood is defined corresponding to the 4[th] annular ring at a distance of 8-10 kpc. The annular rings evolve distinctly in terms of star formation, stellar nucleosynthetic contributions and mixing of interstellar gas. In some simulations we systematically performed radial gas mixing of the interstellar gas and radial gas inflow among the adjacent annular rings in order to understand their influence on chemical evolution. All the simulations are performed with a timestep of one million years (Myr). This timestep serves as the temporal resolution of the present work. The assessments regarding the star formation, the stellar nucleosynthetic contributions to the interstellar medium, the radial gas mixing and the radial gas inflow are made at an interval of 1 Myr. The details of the various simulation parameters are presented in the Table 1 and are discussed in the following.

## 2.1 The accretion model for the galaxy

Several GCE models have been proposed earlier with a wide range of prescriptions for the formation and evolution of the galactic halo, thick and thin discs. We have explored a wide-range of accretion scenarios to understand their implication on the evolution of the galaxy. These models range from a simple two-infall accretion model (Chiappini et al. 1997; Mott et al. 2013) to a recently proposed three-infall accretion model (Micali et al. 2013).

We have used the two-infall accretion model (Matteucci & François 1989; Chiappini et al. 1997, 2001; Mott, et al. 2013) in most of our simulations. The temporal evolution of the total surface mass density, $\sigma(r, t)$ $\left( M_\odot \text{ pc}^{-2} \right)$, of a specific annular grid at a radial distance, $r$, was numerically implemented in the simulations according to the equation 1. The matter, in the form of gas, was gradually added to the galaxy according to the prevailing accretion rate by defining a variable function associated with the gas surface mass density. On the basis of this function the stars were formed as described in the following.

The accretion timescales for the galactic halo and disc in the two-infall accretion scenario are represented by $t_T$ and $t_D$ respectively. The density of the galactic halo is represented as $\sigma_T$. The time, $t_{max}$, is the timescale for the maximum gas accretion on the disc that is assumed to be 1 Gyr. The characteristic timescale is assumed to be 0.8 Gyr for the accretion of the galactic halo.

$$\frac{d\sigma(r,t)}{dt} = A(r)e^{-(t/t_T)} + B(r)\, e^{-(t - t_{max}/t_D)} \tag{1}$$

The quantities, A(r) and B(r), are derived from the normalizing condition of reproducing the current total surface mass density.

$$A(r) = \frac{\sigma_T(r,t)}{t_T(1 - e^{-t/t_T})} \tag{2}$$

$$B(r) = \frac{\sigma(r,t) - \sigma_T(r,t)}{t_D(1 - e^{-(t - t_{max}/t_D)})} \qquad (3)$$

The accretion timescale of the disc is assumed to be a function of galactocentric distance, $r$ (equation 4). This inside-out formation criterion is responsible for the early formation of the inner regions of the disc in comparison to the outer regions (Chiappini et al. 1997, 2001; Micali et al. 2013).

$$t_D(r) = 1.033 \left(\frac{r}{\text{kpc}}\right) - 1.267 \text{ Gyr} \qquad (4)$$

The final acquired current total surface mass density, $\sigma$, subsequent to the complete accretion of disc is assumed to vary from 234 to 6 $M_\odot$ pc$^{-2}$ over the galactic radial extent of 2 to 18 kpc (Rana 1991). The surface mass density of the halo is separately assumed to be 17 $M_\odot$ pc$^{-2}$ from the center to 10 kpc, and thereafter, it is scaled-down with a radial dependence of $r^{-1}$ (Binney & Tremaine 1987; Chiappini et al. 2001; Brusadin et al. 2013) in majority of the simulations (see e.g., Model A in Table 1). This chosen density profile produces a present total surface mass density of 64 $M_\odot$ pc$^{-2}$ in the solar neighborhood at a radial distance of ~8-10 kpc, defined as the solar annular ring. The galactic halo and discs contributions are 17 $M_\odot$ pc$^{-2}$ and 47 $M_\odot$ pc$^{-2}$, respectively. In order to study the variation in the halo surface mass density in the outer regions of the galaxy we ran a simulation (Model B) with a constant surface mass density profile instead of $r^{-1}$ (Table 1). In addition, we ran two simulations (Models I & J) with distinct density profiles for the halo.

In order to understand the evolution of the galactic thick and thin disc, we ran a simulation (Model D) where 20% of the galactic disc mass was assumed to accrete along with the galactic halo. Another simulation (Model H) was run by adopting the three-infall model recently developed by Micali et al. (2013). In this model the halo, thick and thin discs were separately accreted according to the temporal evolution of the density profile defined by the equation 5.

$$\frac{d\sigma(r,t)}{dt} = A(r) \, e^{-(t/t_H)} \; + \; B(r) \, e^{-\left(t - t_{maxH}/t_T\right)} + \; C(r) \, e^{-\left(t - t_{maxT}/t_D\right)} \qquad (5)$$

Here, the first, second and third terms deal with the accretion of halo, thick and thin discs, respectively (Micali et al. 2013). The quantities, A(r), B(r), C(r), are derived from the condition of reproducing the current total surface mass density similar to the scenario for the two-infall accretion model as discussed above. In the case of three-infall accretion, the time of the maximum accretion on the halo and thick disc are represented by $t_{maxH}$ and $t_{maxT}$, respectively. In the present work, $t_{maxH}$ and $t_{maxT}$, are taken to be 0.4 and 1.8 Gyr, respectively. The time-scales, $t_H$ and $t_T$, for the mass accretion of the halo and the thick disc components are taken to be 0.2 and 1.25 Gyr, respectively. The thin disc accreted according to the timescale described in equation 4. The density of the thick disc was assumed to be 24 $M_\odot$ pc$^{-2}$ up to 10 kpc. In the outer regions with lower surface mass densities, the density of the matter accreted during the thick disc phase was assumed to be 40 % of the total disc density (Rana 1991). It should be emphasized that the choices made for the various simulation parameters in the Models D and H are certainly not unique. We do not rule out the other possibilities.

The astronomical observations regarding the accretion of high velocity clouds of low metallicity (~0.1 $Z_\odot$) by the galaxy imposes constraint on the metallicity of the accreting galaxy (see e.g., Alibés et al. 2001; Sahijpal & Gupta 2013). The infalling gas accreting on the galaxy in all eight annular rings was considered to have the metallicity and the stable isotopic abundance of the various elements identical to the metallicity and isotopic abundance acquired by the specific ring at a particular instant of time till the ring acquires a metallicity of 0.1 $Z_\odot$. This corresponds to a numerical value of 0.00143 for the metallicity. Once an annular ring acquires this value the metallicity of the matter accreting afterwards was assumed to be 0.00143, along with the corresponding isotopic abundances of the various stable isotopes.

## 2.2    Star Formation Rate (SFR)

The process of star formation mainly depends upon the amount of gas available for star formation. The simplest form of $SFR$ is proposed by Schmidt (1959) and Kennicut (1998), in which the $SFR$ depends upon the prevailing gas surface mass density, whereas, the more realistic forms also include the dependence on the total surface mass density (Dopita & Ryder 1994). In the present work, the SFR was assumed to depend upon the gas surface mass density, $\sigma_{gas}$, and the total surface mass density, $\sigma$, prevailing within an annular ring at a certain time (Alibés et al. 2001; Matteucci 2003; Sahijpal & Gupta 2013; Spitoni et al. 2015). The radial and the temporal dependence of the normalized $SFR$ (equation 6) was incorporated by defining the star formation efficiency parameters, $\eta(r)$ and $\nu(t)$, respectively. The stellar number distribution function, $G_i(t, m)$, was synthesized at a specific time, $t$, according to the prevailing $SFR$.

$$SFR(r,t) = \nu(t)\, \eta(r)\, \frac{\sigma^n(r,t)\sigma_{gas}^m(r,t)}{\sigma^{n+m-1}(r,t)} \; M_\odot \; pc^{-2} \; Myr^{-1} \qquad (6)$$

Here, $n$ and $m$ are the simulation parameters with the assumed values of 1/3 and 5/3, respectively (Dopita & Ryder 1994; Alibés et al. 2001; Sahijpal & Gupta 2013). We find a strong dependence of the GCE models on the star formation efficiency parameters. In order to explain the observed elemental and isotopic abundance distribution trends across the galaxy we optimized our simulations with the radial dependence of the star formation efficiency parameter, $\eta(r)$ in the range of 4 to 0.2 from the 1st to 8th annular rings during the entire temporal evolution of the galaxy in the case of Models A, B, D-H (Table 1). In order of the annular ring number from the galactic center, these efficiency values are 4, 3, 2, 1, 0.7, 0.45, 0.3 and 0.2. The value of unity is chosen for the annular ring corresponding to the solar neighborhood as a normalization. In terms of the temporal dependence of the star formation efficiency parameter, $\nu(t)$, the SFR during the initial 1 Gyr that is characterized mostly by halo phase was assumed to be 2. This parameter was assumed to be unity for the evolution subsequent to 1 Gyr.

In the case of Models C, I and J (Table 1), the radial dependent star formation efficiency parameter, $\eta(r)$, was assumed to be unity over the entire galaxy. The time dependent star formation efficiency parameter, $\nu(t)$, in these simulations were assumed to be identical to the remaining simulations. The comparison of the deduced elemental abundance distribution of the Model C with other models provides an understanding about the nature of star formation efficiencies. Further, the Models I and J that consider radial inflow of gas are capable of reproducing most of the elemental observed patterns even without considering the dependence on the parameter, $\eta(r)$. In the case of three-infall accretion model, the parameter, $\eta(r)$, was assumed to be identical to the Model A, whereas, the parameter, $\nu(t)$, was assumed to be 2, 4 and 1, for the halo, thick and thin disc phases, respectively. These values are distinct from the parametric values originally proposed by Micali et al. (2013). We found the best optimization for this set of values.

## 2.3    The stellar initial mass function

Stars are born in clusters with a mass distribution that exhibit a power law distribution (Salpeter 1955; Pagel 1997; Alibés et al. 2001; Matteucci 2003; Sahijpal & Gupta 2013). The stellar IMF defines the number of stars of a particular mass that are formed in a stellar cluster. The power law distribution according to equation 7 is generally considered for stellar IMF.

$$\phi(m) = A\, m^{-(1+x)} \tag{7}$$

Here, A is a normalization constant and $x$ is the power law exponent (Kroupa 1998; Scalo 1998; Matteucci 2003). The elemental evolution history of the galaxy can be reproduced efficiently using multiple slope in the IMF (Matteucci 2014; Sahijpal & Gupta 2013). The slope is almost flat for the low-mass stars and becomes steep in the case of massive stars. In the present work, we have adopted a formulation distinct from the one used by Sahijpal & Gupta (2013) and Sahijpal (2014) in defining the stellar IMF. Instead of a discrete mass stellar IMF with three distinct power law exponents (Scalo 1998; Pagel 1997; Matteucci 2003), we have used a discrete IMF with two distinct power exponents over the entire mass range. The value of $x$ is chosen to be 0 in the mass range 0.1-1 $M_\odot$ and 1.7 in the mass range of 1-100 $M_\odot$ for all the simulations. It is possible to run simulations with other choices of the power exponents. However, we avoid parametrization of the IMF power exponents. A normalized discrete IMF based on the equation 7 is used in all the simulations. Stars with discrete masses, $m$, of 0.1, 0.4, 0.8, 1, 1.25, 1.75, 2.5, 3-8 and 11-100 $M_\odot$ in the IMF were chosen to adequately cover the entire mass range. The simulation stars in the mass range 3-100 $M_\odot$ have integer masses except for the gap between 9-10 $M_\odot$. The stellar nucleosynthetic yields within this small mass interval are not available and the possible evolutionary fate of the stars remains indefinite because of uncertain mass losses the stars experience. However, in order to appropriately cover the most populous stellar low-mass range of 0.1-3 $M_\odot$ in the IMF, we considered numerous stars of distinct masses. One of the major reasons for the specific choice of these stellar masses is the availability of the stellar nucleosynthetic yields of the AGB stars corresponding to these masses (see e.g., Karakas & Lattanzio 2007). In addition, the stellar population evolution of the 0.8, 1 and 1.25 $M_\odot$ was collectively monitored over the galactic evolution to understand the G-dwarf metallicity distribution.

In order to obtain a normalized IMF defined by equation 7 with discrete stellar masses it is essential to either consider stars with a unit solar mass interval over the entire IMF as in the case of the stellar upper-mass range of 3-100 $M_\odot$ or obtain a proper normalization with a comparatively low stellar mass contributions from each star belonging to the low-mass range that has less than a unit solar mass interval. We followed the second approach by taking proper precautions to avoid overestimating the mass contribution from the stellar low-mass range to the IMF in comparison to the stellar upper-mass range. This was achieved by redetermining the total number of stars in the IMF corresponding to each stellar mass in the low-mass range. The total number of 1 $M_\odot$ stars, according to the IMF defined by equation 7 with a power exponent of 1.7, were re-distributed proportionally among the 1 and 1.25 $M_\odot$ stars according to their weightage defined by the IMF with the power exponent of 1.7. This circumvented the overestimation in the total number of these two stars with less than a unit mass difference. In a similar manner, the number of stars corresponding to the stellar masses, 1.75 and 2.5 $M_\odot$ were determined by estimating the total number of 2 $M_\odot$ stars according to the IMF defined by equation 7 with a power exponent of 1.7, and proportionally distributing them according to the IMF. Finally, the number of 0.1, 0.4 and 0.8 $M_\odot$ stars were estimated on the basis of the renormalized number of 1 $M_\odot$ stars as estimated above and the power exponent of 0 in the low-mass range of 0.1-1 $M_\odot$. The final normalized IMF obtained in the above manner has a ~0.45 and 0.55 mass fraction contributions from the stars in the mass ranges, 0.1 - 0.8 $M_\odot$ and 1-100 $M_\odot$, respectively, towards the integrated IMF.

Subsequent to each simulation timestep of 1 Myr, the amount of gas mass available for star formation is estimated in each annular ring according to *SFR* and the stellar number distribution function, $G_i(t, m)$, is synthesized for the stars according to the normalized IMF and the metallicity prevailing in the ring at the specific time, $t$.

## 2.4    Stellar Nucleosynthetic yields

The simulation stars in the mass range 0.1-100 $M_\odot$ are evolved over lifespans that depend upon the stellar metallicity and mass (Pagel 1997). At a given time, $T$, an independent assessment is made for each annular ring, $i$, for all the previously synthesized stellar number distribution functions, $G_i(t, m)$, such that t < T, to ascertain all the stars that would contribute their nucleosynthetic yields. The stellar ejecta from these stars are homogenized instantaneously within the entire annular ring, thereby, enriching the metallicity of the ring. We have adopted the stellar nucleosynthetic yields of low and intermediate mass AGB stars (0.8-8 $M_\odot$) by Karakas & Lattanzio (2007).

We used the stellar nucleosynthetic yields of the massive stars (> 11 $M_\odot$) obtained by Woosley & Weaver (1995). The stars in the mass range 34-100 $M_\odot$ evolve through Wolf-Rayet stage followed by SN I b/c (Sahijpal & Gupta 2009), whereas, the stars in the mass range 11-33 $M_\odot$ explode as SN II supernova. A recent comparison study (Sahijpal 2014) among the GCE models based on the stellar nucleosynthetic yields of the massive stars obtained by two distinct research groups (Woosley & Weaver 1995; Chieffi & Limongi 2004, 2013) indicate that it is not possible to obtain all the observed elemental evolutionary trends in a self-consistent manner. The stellar nucleosynthetic yields obtained from the various available stellar models are used to interpolate and/or extrapolate the unavailable yields over the entire range of masses and metallicities as encompassed during the simulations. Unlike the previous work (Sahijpal & Gupta 2013), we have not reduced the stellar nucleosynthetic yields of the iron-peaked nuclides in the present work. We have used the supernova model A yields obtained by Woosley & Weaver (1995) for the massive stars. It should be noted that Timmes, Woosley & Weaver (1995) have suggested that the yields of the iron-peaked nuclides should be reduced by a factor of 2 for the massive stars.

Single degenerate stars (SD) in the binary systems with the mass in the range of 3–8 M$_\odot$ and 11–16 M$_\odot$ can evolve to supernova SN Ia. SN Ia is considered to be the main contributor of iron in the galaxy. We numerically synthesized a binary stellar population, $B_i(t, m)$ {$m \subset$ m}, from the stellar number distribution function, $G_i(t, m)$, at the time of formation of the simulation stars by assuming a number fraction, $f$, of the stars that evolves into binary systems with a definite progenitor of SN Ia. The mass, $m$, is in the range of 3–8 M$_\odot$ and 11–16 M$_\odot$. The remaining stellar number fraction in the specific mass range evolves as single stars either as intermediate mass AGB stars or massive stars. The parameter, $f$, is treated as a simulation parameter to obtain the value of 0.0143 for the solar metallicity and [Fe/H] = 0 in the solar neighborhood at the time of formation of the solar system around 4.56 Gyr ago. This parameter has a typical value in the range of 0.02-0.025 in our simulations. The binary stellar population, $B_i(t, m)$, generated at a specific time, $t$, within an annular ring, $i$, is integrated over the specific mass range to obtain a *single pool of binary stars*. The delay time in this binary pool of stars for SN Ia explosion is randomly assigned according to a probability distribution function based on a delay time distribution (DTD). We adopted the DTD proposed by Matteucci & Recchi (2001) [MR01] for the single degenerate (SD) models in most of our simulations as it has been considered as one of the most acceptable SN Ia formulation (Matteucci et al. 2009). The SN Ia delay time [MR01], represented by *Log(t)* in the range of 7.55 to 10.05, is binned into 26 intervals of 0.1 bin size each. A probability distribution function based on random numbers is synthesized for the complete temporal range. A random number is generated for each SN Ia candidate from a single pool of binary stars to determine the time of explosion. The DTD ensures a rapid initiation of SN Ia within the initial 35 Myr after a single burst of star formation. Approximately, 0.1 fractional contribution of prompt SN Ia are expected within the initial <100 Myr. The remaining SN Ia explode with longer time delays.

As an alternative, we ran a simulation (Model E) with the DTD based on double degenerate (DD) binary system (Greggio 2005 [G05]). An identical procedure is adopted to synthesize a probability distribution function based on random numbers. We can perform simulations for any choice of DTD. Finally, the supernovae type Ia nucleosynthetic yields obtained by Iwamoto et al. (1999) are used in the simulations.

## 2.5 Radial gas mixing and gas inflow

As discussed in the previous section, the redistribution of matter in terms of stellar migration, gas mixing and gas inflow occurs in the disc because of dynamical evolution of the galaxy, the spiral structure of the disc and the infalling gas on the disc. The radial migration of matter along the disc can substantially influence the chemical evolution of galaxy. The majority of the simulations in the present work have been performed with an independent galactic evolution of the eight annular rings. However, we have developed a procedure to numerically simulate (Table 1) the radial gas mixing among the adjacent annular rings on the basis of the average observed radial gas velocities of ~1 km s$^{-1}$, or identically, ~1 kpc Gyr$^{-1}$ (Portinari & Chiosi 2000; Schönrich & Binney 2009; Spitoni & Matteucci 2011; Spitoni et al. 2015). Due to the adopted nature of the stellar number distribution function, $G_i(t, m)$, it is also possible to consider stellar radial migration (Minchev et al. 2013, 2014) by systematically re-distributing the stars across the various annular rings, specially, the $< 2 \, M_\odot$ stars that can drift substantially during their lifetimes with the observed typical radial velocities of ~1 km s$^{-1}$. However, we have restricted our present work to incorporate the influence of radial gas mixing and inflow on GCE. As mentioned earlier, the influence of the stellar radial migration on GCE will be considered elsewhere in details. However, we will qualitatively discuss the influence of stellar migration in our present work.

We assumed a flat disc geometry for the radial mixing of gas among the annular rings. We have performed the partial mixing of the isotopic inventories of the interstellar gas in a specific annular ring, $i$, with the isotopic inventories in the adjacent annular rings, $i\text{-}1$ and $i\text{+}1$, separately, at every timestep of 1 Myr. During this simulation timestep the interstellar gas would radially move over a distance of approximately one parsec if we assume an average observed radial gas velocity of ~1 km s$^{-1}$. Thus, in terms of the fractional annular area, approximately 0.06 % of the annular ring gas will radially move across the adjacent annular ring during 1 Myr. The simulation, 'Model F', is performed with 0.06 % gas mixing per one million years among the adjacent annular rings (Table 1). We have also performed a simulation, 'Model G', with 0.23 % gas mixing per one million years. This corresponds to a scenario of radial gas mixing with an assumed radial gas velocity of 4 km s$^{-1}$.

We have implemented the radial gas inflow in the Models I and J. The inflow velocities are varied with galactocentric radius according to the suggestions made by Spitoni & Matteucci (2011) and Mott et al. (2013). The radial inflow velocities are assumed to be, 0, 0, 0.5, 1, 2, 3 and 3.7 km s$^{-1}$ at the outer edge of the annular rings, listed in the sequence from the 1$^{st}$ to 7$^{th}$ ring. The inflow velocity reduces as we move inwards in the galaxy. We assumed no net inflow of gas in the 8$^{th}$ annular ring. The radial inflow of gas was numerically executed by considering a percentage fraction, $\chi_i$, of the gas that inflows from the i+1$^{th}$ annular ring to the i$^{th}$ ring during the simulation timestep of 1 Myr. If we assume that the entire interstellar gas fraction within an annular ring participates in the radial inflow, the percentage area of the annular rings involved in the radial mass inflow across a ring during 1 Myr with the chosen radial inflow velocities of 0, 0, 0.5, 1, 2, 3 and 3.7 km s$^{-1}$ would be 0, 0, 0.02, 0.05, 0.09, 0.14 and 0.17 percentage, respectively, from the 1$^{st}$ to 7$^{th}$ ring. This corresponds to a substantial inflow of gas causing a strong depletion of gas as well as the total surface matter density in the outer annular rings at the cost of matter accumulation in the inner rings. In order to circumvent this excessive inflow of gas towards the center we have assumed 0.1 mass fraction of the total available gas fraction to be actively involved in the inflow. Thus, we used the values of 0, 0, 0.002, 0.005, 0.009, 0.014 and 0.017 percentage for $\chi_i$, from the 1$^{st}$ to 7$^{th}$ ring. The assumed radial profiles of the halo surface mass density in the Models I and J are presented in Table 1. The radial stellar formation efficiency parameter, $\eta(r)$, in these simulations is assumed to be unity over the entire galaxy. As discussed latter in the section 4, the radial gas inflow provides a natural mechanism to obtain the observed elemental abundance gradients without adopting a preferred set of stellar efficiency parameters. Finally, the choice of the simulation parameters made in the present work is certainly not restricted (Mott et al. 2013). We have also simulated a Model J in order to incorporate the radial gas inflow superimposed over the radial gas mixing. The mixing was performed in a manner identical to the Model F with a mixing velocity of 1 km s$^{-1}$. We have not specifically performed the simulation with radial gas outflows that are mainly caused by the energetic winds from supernovae and other massive stars. The radial gas mixing models takes care of this aspect to an extent.

## 3.      Results

The major aim of the present work was to develop a Monte Carlo based GCE model for the Milky Way galaxy. We ran a large number of simulations to optimize the various simulation parameters based upon the observational constraints. A selective set of simulation results for our three best Models A, H and J are presented in the Fig. 1-10 and Table 1. We have avoided the detailed discussion and the presentation of the simulation results for the Models B-G and I in the main text. The results of these models can be found in the Fig. 11-31 in the supplementary data. However, we summarize the major findings of these models in the main text. We are presenting a comprehensive account of the deduced evolution of the SFR, the [Fe/H], the metallicity $Z$ and the stellar birth rate for the various annular rings of 2 kpc width each, starting from 2 to 18 kpc from the galactic center. The deduced temporal evolution of the total surface mass density (stars+interstellar gas), $\sigma_{\text{Total}}$, the surface mass density of stars $\sigma_{\text{stars}}$ and the surface mass densities of the stellar remnants (Neutron stars + Black holes) and white dwarfs are also presented for the various annular rings. These simulation results for the models A, H and J are presented in the Fig. 1-3.

Since, a lot of computation is involved in numerically simulating the galaxy, we have performed detailed mass-balance calculations at each temporal step of 1 Myr in a simulation for each annular ring. We are graphically presenting here the propagation of the cumulative error that is defined according to the relation, $|\{(\sigma_{\text{Total}} - \sigma_{\text{gas}}) - (\sigma_{\text{stars}} + \sigma_{\text{White dwarf}} + \sigma_{\text{Neutron star + Black hole}})\} \times \text{Ring area}|$ (in $M_{\odot}$) for the various annular rings. The assessment of the cumulative error is extremely helpful to monitor the correctness of the numerical code. The various densities involved in the calculation of the cumulative error are estimated independently. The error estimation involves all the physical processes resulting in the accretional growth of the galaxy, the episodes of star formation and the stellar nucleosynthetic contributions of the evolved stars. This error should be strictly zero. However, our simulations infer a significantly low cumulative error that in terms of the time-integrated cumulative error for the entire galaxy from 2-18 kpc turns out to be $< 0.15$ $M_{\odot}$. This can be ignored at the scale of the entire galaxy and even at the level of an individual annular ring.

The normalized elemental abundance distribution, generally refereed as the abundance gradients, of nine selective elements across the galaxy at seven distinct temporal epochs of evolution for the Models A, H and J are presented in the Fig. 4-6. We have also presented the observational data based on the Cepheid variables, dwarf stars, red giants, PNe and OB stars. The normalization in all cases is done with respect to the solar abundance at 9 Gyr at the solar annular ring between 8-10 kpc. In the recent times the availability of the observation data from *RAVE, GAIA, APOGEE, SEGUE, LAMOST, GCS*, etc., has substantially helped to constraint several parameters related with GCE models. The red giants being cool and highly luminous provide a good observational sample due to their wide range of lifespans and chemical composition (e.g., Cescutti et al. 2007; Colavitti et al. 2009; Miglio et al. 2013; Anders et al.

2014). Further, we have taken the observational data of the Cepheid variables. A high-resolution spectral data has been obtained to determine the abundance of light, alpha and heavy elements over the galactic distance of 7-12 kpc. Boeche et al. (2013) analyzed a sample of dwarf stars from *RAVE* and *GCS* datasets to provide the gradients of Mg, Al, Si and Fe. Cunha et al. (2016) obtained the metallicity gradient of red giant star over the galactocentric distance of 8-15 kpc from the *APOGEE* survey. Abundance gradient data for the inner and the outer regions of the galaxy is provided by Martin et al. (2015) and Andrievsky et al (2016). Andrievsky et al. (2002a, b, c, 2004) determined the abundance gradients from the Cepheid variables. Luck et al. (2003) obtained the abundance gradients in the outer parts of the galaxy from Cepheid variables. Daflon & Cunha (2004) investigated the OB stars, OCs and H II regions over the galactocentric distance of 4.7-13.2 kpc. Carney et al. (2005) studied a sample of red giant stars. Yong, Carney & de Almeida (2005) and Yong et al. (2006) studied stars in OCs and Cepheid variables in the outer regions of the galaxy. Other observational data is taken from Simpson et al 1995; Afflerbach, Churchwell & Werner 1997; Twarog, Ashman & Anthony-Twarog 1997; Carraro, Ng & Portinari 1998; Gummersbach et al. 1998 and Chen, Hou & Wang 2003. A detailed reference list of the observational data is provided in the Fig. 4-6 for the various evolved stars.

The temporal evolution of the normalized [X/Fe] vs. [Fe/H] for the Models A, H and J is presented in the Fig. 7-9. Here, X, represents the various low and intermediate mass elements. The normalization in all cases is done with respect to the solar abundance at 9 Gyr in the solar annular ring between 8-10 kpc. The observational data for comparison with our deduced trends is also presented in the figures. The observational data for [Ne/Fe] vs [Fe/H] is not available as it is not easy to observe Ne lines. However, it should follow same trend as the other α-nuclides (Goswami & Prantzos 2000).

Finally, the G-Dwarf metallicity distribution of stars at the present epoch for all the annular rings is presented the Fig. 10. We are including the stars in the mass range 0.8-1.25 $M_{\odot}$ to determine the G-Dwarf metallicity distribution. The predictions are compared with the observational data obtained by Jørgensen (2000) and Holmberg, Nordström & Andersen (2007).

## 4.      Discussion

We have successfully developed the numerical simulations with majority of the features related with the galactic chemical evolution (GCE) of the Milky Way galaxy in spite of the distinct numerical approach adopted in comparison to the earlier GCE models. The advantage of the present adopted approach is that the GCE simulations can be performed at a discrete level of an evolving ensemble of stars without theoretically going through the complex analytical formulations of the models based on the integro-differential equations (Pagel 1997; Matteucci 2003). The simplicity of the approach makes it easier to implement compared to the conventional GCE models. In our GCE models we have incorporated distinct galactic accretion scenarios, the role of the SN Ia delay time distribution function (DTD), the influence of the radial gas mixing and the radial gas inflow on the evolution of the galaxy. All these features are implemented by defining a stellar number distribution function, $G_i(t, m)$, for an evolving ensemble of stars. In general, the major assumptions related with the adopted gas accretion scenarios, the *SFR* and the stellar *IMF* enable us to successfully predict the majority of the observed elemental abundance evolutionary trends across the galaxy. We discuss some of the salient features of our simulation results in the following. However, it should be mentioned that we have tried to explore a range of parametric space associated with GCE. We cannot rule out the possibility of alternative choices of the various simulation parameters.

### 4.1 Star formation rates

The assumed radial dependence of the star formation efficiency parameter, $\eta(r)$ (Table 1; Fig. 1-3) is essential in most of the simulations to explain the observed elemental abundance gradients (Fig. 4-6), the radial spreads in the temporal evolution of [Fe/H] (Fig. 1b-3b) and the metallicity, Z (Fig. 1c-3c) across the galaxy except in the case of the Model J, where we have considered the radial gas inflow. The radial gas inflow provides a natural mechanism to explain the observed elemental abundance gradients and the radial spreads in the temporal evolution of [Fe/H] and the metallicity, Z, without considering the radial dependent star formation efficiency parameter. The Model C (Table 1) in which neither of these two dependencies are considered, the observed elemental abundance gradients cannot be achieved. Further, in terms of the [X/Fe] (e.g., [O/Fe]) vs. [Fe/H] evolution, the Model C (Fig. 16, supp. data) exhibits a narrow range of possible elemental evolutionary trends across the galaxy compared to the Model A (Fig. 7). It is not possible to explain the observed trends in the [X/Fe] vs. [Fe/H] evolution for the entire galaxy in the case of Model C even if we anticipate radial stellar migration.

The requirement of an enhanced *SFR* during the galactic halo epoch is based on the elemental abundance evolutionary trends. This dictates the use of a high value of $\nu(t)$, the temporal stellar formation rate efficiency parameter. We have used a constant value of 1 for this parameter during the disc epoch and a value of 2 during the halo epoch in all the simulations with two-infall accretion scenario. In general, the *SFR* exhibits a rapid enhancement during the initial 1 Gyr (Fig. 1a-3a). Since, a substantial amount of the initially accreted gas in the galaxy is consumed during the initial bursts of star formation, the *SFR* rate declines subsequent to the initial 1 Gyr. The dip in the *SFR* in the case of two-infall accretion scenario around ~1 Gyr represents the transition from the galactic halo to disc epoch.

In the case of Model H, with an assumed three-infall accretional growth, we optimized the simulations with a value of 4 for $\nu(t)$ in the thick disc phase. This is quite distinct from the choice of 10 made for this parameter by Micali et al. (2013) for the solar neighborhood. The merger of protogalaxies with the accreting Milky Way galaxy is considered to be the reason for the high star formation efficiency, $\nu(t)$, during the thick disc phase. Although the set of parametric values used in our simulations are certainly not unique yet the adopted choice enabled us to explain the majority of the features associated with the normalized elemental abundance evolution (Fig. 4-6).

In the three-infall accretion scenario, the halo forms in almost half of the time that it takes to form in the case of two-infall accretion scenario. This is followed by distinct thick and thin disc epochs. Compared to the Models A & D, with two-infall accretion scenario (Fig. 7 & 19), in the case of Model H, with three-infall accretion scenario, the observed spreads in the [X/Fe] vs. [Fe/H] evolutionary trends (Fig. 8) are explained in a better manner, specially, for the thick disc stars. The match between the observed and the deduced elemental evolutionary trends can be improved further in the case of Model H if we increase the value of $\nu(t)$ to more than 4 as in the case of the models proposed by Micali et al. (2013). However, we avoid the use of a higher value as the net stellar efficiency parameter, $\nu(t) \times \eta(r)$, in the inner regions of the galaxy increases up to a value of 40 that is not conceivable without any theoretical justification. In general, the three-infall accretion model (Micali et al. 2013) seems to be a viable scenario for the galaxy accretion and evolution compared to the two-infall accretion model. We anticipate that the incorporation of radial stellar migration, radial gas mixing and radial gas inflow would further improve the match between the predicted and the observed elemental abundance trends for the three-infall accretion models.

The numerically deduced stellar birth rates for all the annular rings of the galaxy are presented in the Fig. 1d-3d. The stellar birth rates in the inner most annular ring exhibit a gradual decline subsequent to the initial ~4 Gyr. This is due to the rapid mass accretion and star formation during the initial stages.

## 4.2  Accretion of the galaxy and the temporal evolution of the distribution of matter

The deduced total surface mass density of the galaxy following the two-infall as well as the three-infall accretional scenarios is presented in Fig. 1e-3e. The initial 1 Gyr defines the most rapid phase of the accretion of galaxy. The distribution of the galactic mass in terms of the stellar surface mass density (Fig. 1f-3f)), the stellar remnant mass densities corresponding to the neutron stars + black holes (Fig. 1h-3h) and the white dwarfs (Fig. 1i-3i) exhibit evolutionary trends that are in general accordance with the accretional growth of the galaxy and the star formation rates. In general, the amount of the galactic mass that gets eventually entrapped by the white dwarfs is >10 times more than the mass that gets entrapped collectively by neutron stars and black holes.

In the case of Models, I & J with radial gas inflows, the gas surface mass density is presented in Fig. 2g & 3g to ensure that the radial gas inflow does not create substantial reduction of the matter density in the outer regions of the galaxy at the cost of matter accumulation in the inner regions. As discussed in the section 2, we have assumed 0.1 mass fraction participation of the total available gas in the annular rings towards the radial gas inflow. We have not precisely parametrized this mass fraction as the present predicted total surface mass density in the outer regions of the galaxy lies with a factor of 2 in the Models I & J (Fig. 2e & 3e) as compared to the Model A (Fig. 1e) with no gas inflow.

## 4.3 Supernova rates

The deduced supernovae (SN Ib/c and II) rates (Fig. 1k-3k and 1l-3l), in general, follow the temporal evolution trends identical to the *SFR* (Fig. 1a-3a) as the initiation of these supernovae explosions occur within few million years after the formation of stars. The stellar nucleosynthetic contributions of supernovae (SN Ib/c and II) dominate the elemental abundance evolution during the galactic halo phase (Fig. 7-9). The sharp dips in the deduced supernovae (SN Ib/c and II) rates around 1 Gyr marks a resemblance with the deduced *SFR*.

The supernova (SN Ia) rate does not follow the SFR trends in time as the SN Ia progenitors are long-lived compared to massive stars. We have used two distinct DTD [MR01 & G05] (Table 1) in the present work. The adopted approach of synthesizing the SN Ia population based on random numbers produces fluctuating trends in the deduced SN Ia rates. SN Ia contribution commence well within the initial 1 Gyr (Fig. 1j-3j) as discussed in the section 2. There is an observed reduction in the rates around 1 Gyr. However, the dip around 1 Gyr is less pronounced compared to the dip in the other supernovae types. The dip almost disappears in the case of the three-infall accretion model (Fig. 2j). Our results almost agree with the SN Ia rates deduced by Micali et. al (2013).

There is a marked difference in the deduced SN Ia rates for the models with the two DTD [MR01 & G05]. The maxima in the SN Ia rates shifts from the initial 3-4 Gyr in the case of Model A (Fig. 1j) based on [MR01] to ~2 Gyr in Model E (Fig. 20j; supp. data) based on [G05]. This is substantially reflected in the deduced evolutionary trends in [X/Fe] vs. [Fe/H] in Fig. 7 & 22), where the former model indicates a better match with the observed elemental abundance trends. The differences in the two models are due to the fact that the DTD based on the double degenerate (DD) progenitor model (Greggio 2005) produces a smaller population of SN Ia progenitors with more than 1 Gyr delayed time compared to the single degenerate (SD) model (Matteucci & Recchi 2001). This results in an early nucleosynthetic contribution from SN Ia to the galactic elemental inventories in the case of [G05] based DTD.

### 4.4 The temporal evolution of the Metallicity (Z) and [Fe/H]

We have assumed a metallicity value of 0.0143 (Asplund et al. 2009) and a value of ~0 for [Fe/H] in the solar neighborhood at the time of formation of the solar system around 4.56 billion years ago. The spreads in the deduced temporal evolution of [Fe/H] and metallicity (Fig. 1b-3b & 1c-3c) across the galaxy impose stringent constraints on the basic choice of the simulation parameters. As mentioned earlier, there is a substantial reduction in the inferred spread across the galaxy in [Fe/H] and metallicity in the case of Model C (Fig. 14b, c; supp. data) in comparison to the Model A (Fig. 1b-c). The star formation efficiency parameter, '$\eta(r)$' is assumed to be constant in the former case. This reduces the possibility of the Model C to be a viable scenario for GCE.

Further, in comparison to the Model A (Fig. 1b, 1c), the deduced spreads in [Fe/H] and Z evolution across the galaxy reduce in the case of Models F and G (Fig. 23b, 26b, 23c, 26c; supp. data) corresponding to the scenarios where the radial gas mixing among the adjacent annular rings is considered at a rate of 0.06 % and 0.23 % of ring mass mixing per million years, respectively. The Models F and G corresponds to the gas mixing velocities of 1 and 4 km s$^{-1}$, respectively. The substantial reduction in the spreads of [Fe/H] and Z evolution in the latter case impose stringent constraints on the magnitude of the velocity associated with gas mixing among the two adjacent annular rings to be < 4 km s$^{-1}$. The deduced spreads in [Fe/H] and Z evolution across the galaxy are small in the case of Model J (Fig. 3b, 3c) that incorporates both the radial gas inflow and radial gas mixing compared to the Model I (Fig. 29b, 29c; supp. data) that does not incorporate radial gas mixing. However, the differences are not significant to rule out the possibility of the scenario corresponding to the Model J. It should be noted that both models reproduce spreads in the elemental abundance gradients across the galaxy (Fig. 6, 30) even without radial dependent star formation efficiency parameter.

### 4.5 Elemental abundance gradients

Astronomical observations indicate the presence of radial elemental abundance gradients along the galactic disc. Our model predictions for the normalized abundance gradients of the elements, C, N, O, Mg, Si, Ca, Ti, Fe and Zn, are compared with the observational data (Fig. 4-6) that includes the elemental abundance distribution of stars during various evolutionary stages. These various evolutionary stages are mentioned in the Fig. 4-6. The elemental abundance gradients are shown at seven different time-epochs of the galactic evolution, viz., 0.5, 1, 2, 4, 9, 11 and 13.5 Gyr, from the initiation of the formation of the galaxy. The predicted gradients are steep in the inner regions and the steepness reduces in the outer regions. The inside-out formation criterion of the galaxy is responsible for the early formation of inner regions. The high *SFR* during the earliest times (1 Gyr) in the inner regions resulted in the fast elemental enrichment of the interstellar medium. Hence, the elemental abundances are high in the inner regions.

In majority of the simulations, the steepening of the gradients increases with time in the inner regions but the outer regions do not change much (Chiappini et al. 2001). The steepness reduces substantially around 6-7 kpc over the timescale of 9 Gyr There is no further change in the gradients beyond this time epoch. As discussed earlier in the section 4.1, the Model C (Table 1) infers no substantial elemental abundance gradients beyond solar neighborhood due to the assumed constant star formation rate efficiency. The gradients in the Model D and E (Fig. 18, 21; supp. data) are almost identical to the gradients estimated for the Model A (Fig. 4) except for the differences during the initial 2 Gyr evolution in the case of Model E. Thus, the extent of accretion of mass during the halo phase (see e.g., Model D) does not substantially influence the deduced elemental abundance gradients. The Model E based on the [G05] SN Ia DTD produces distinct gradients from the Model A based on the [MR01] SN Ia DTD during the initial ~2 Gyr. The gradients are identical afterwards.

In comparison to the Model A (Fig. 4), the Model F (Fig. 24; supp. data), with radial gas mixing at the rate of 0.06 % ring mass per one million years among the adjacent annular rings produces smaller gradients due to gas mixing that homogenizes the interstellar medium among two adjacent annular rings. The Model G (Fig. 27; supp. data), with radial gas mixing at the rate of 0.23 % ring mass per one million years, exhibit abundance gradient trends with almost zero slope due to excessive homogenization. Along with the deduced small spreads in [Fe/H] and metallicity as discussed in section 4.4, the deduced elemental abundance gradients make the scenario with the Model G nonviable in terms of the evolution of the galaxy. Thus, we impose a stringent constraint on the average velocity of the radial gas flow for elemental (isotopic) homogenization to be $\leq 1$ km s$^{-1}$. This is consistent with the previous works (Portinari & Chiosi 2000; Spitoni et al. 2015).

In the case of Model H (Fig. 5) with three-infall accretion scenario, the gradients have almost identical slope as that for the Model A during the late stages of evolution (Fig. 4). However, in the Model H, we are able to achieve the inversion in the gradients around 2 Gyr. The inversion feature is also present in the case of Models I and J (Fig. 30, 6) where the radial gas inflow among the annular rings is incorporated. The inversion in the gradients have been predicted by Cresci et al. (2010). It has been recently reproduced analytically in the GCE models by Mott et al. (2013). As an important finding of the present work, we observe inversion for all the element, whereas, in literature inversion is not generally collectively shown for all the elements. In general, the Models H and J seem to represent more viable and realistic scenarios for the accretion and evolution of the galaxy in terms of the deduced elemental abundance gradients. In comparison to the Model I, the Model J, with radial gas mixing, performs better in terms of explaining the elemental abundance gradients in the outer galaxy. To summarize, the models with three-infall accretion model, radial gas inflow and radial gas mixing perform better in majority of GCE aspects related with observations.

Stellar migration also significantly influences the GCE trends (see e.g., Minchev et. al 2013, 2014) as it produces a wide spread in the abundance distribution. Majority of the migrated stars restore the kinematical and dynamical properties of their birth place, e.g., the birth places of the sun is predicted at 4.4-7 kpc. Migration of stars has also been proposed as the reason for the thickening of the disc and the formation of thick disc. We have not incorporated the influence of stellar migration in the present work. It would be considered elsewhere.

## 4.6 The [X/Fe] vs. [Fe/H] evolution

The [X/Fe] vs. [Fe/H] evolutionary trends (Fig. 7-9) are widely used to study the chemical enrichment history of the interstellar medium. We have studied several elements (C, N, O, Ne, Mg, Al, Si, S, Ca, Ti, Ni and Zn). The [X/Fe] vs. [Fe/H] evolutionary trends help to unfold the different timescales for the evolution of distinct elemental ingredients of the galaxy. The [$\alpha$/Fe] vs. [Fe/H] evolution is specifically important in this aspect because of the differences in the timescales of the nucleosynthesis of the $\alpha$-nuclides and iron. Iron is mostly produced by SN Ia and $\alpha$-nuclides come from SN II and SN Ib/c. Since, the timescale for the evolution of massive stars is much shorter than SN Ia timescale, the [$\alpha$/Fe] vs. [Fe/H] gradually reduces (Fig. 7-9). The presence of the knee feature in the evolutionary trends marks the transition of the galaxy from the halo to disc phase. Our normalized elemental evolution predictions do not exactly fit well with the observational data in all the cases. We have strong disagreements in the case of evolution of [Al/Fe], [Ti/Fe], [Ni/Fe] and [Zn/Fe], initiating right from the halo phase. The most probable reason for this disagreement could be due to the uncertainties in the stellar nucleosynthetic yield of these elements from the massive stars, specially, subsequent to the recent revisions in the solar metallicity. The stellar nucleosynthetic yields used in the present work are based on the stellar models with the initial abundances derived from the pre-revised solar elemental abundances. The mass fraction contributions of various elements have been substantially modified subsequent to the revision in the solar metallicity. The disagreements between the observed and the GCE predicted elemental abundance distributions have been discussed in details for the solar neighbourhood by Sahijpal & Gupta (2013).

Further, the observational spread in the elemental abundance distribution cannot be exclusively explained by the spreads in the deduced elemental abundance evolution trends over the entire galaxy, specifically, the thick disc stars (Fig. 7-9). The Model H (Fig. 8), with the three-infall accretion growth of the galaxy infers a better match in terms of the spread in the observed elemental abundance evolution of the thick disc stars. The incorporation of radial stellar migration in this model would be able to explain the majority of the observed features of the galaxy. In addition, the incorporation of the radial gas inflow and radial gas mixing as in the case of Models I & J would also explain the inversion in the initial elemental abundance gradients. The inversion in the elemental abundance gradients is also reflected in the elemental abundance evolution of these models (Fig. 9, 31).

### 4.7 G-Dwarf metallicity distribution

The G-Dwarf metallicity distribution provides a major constraint on GCE models in terms of determining the evolution of the population of low-mass stars in the mass range ~0.8-1.2 $M_\odot$. The G-Dwarf metallicity distribution for each annular ring for six models is presented in Fig. 10. In comparison to the outer annular rings, the inner rings exhibit a wide spread in [Fe/H], thereby, demonstrating a more evolved inner region in terms of star formation and evolution. Among the various models, the Models A, D, H and J yield better results for the stars in the solar neighbourhood as their [Fe/H] distribution match with the observed distribution. The radial gas mixing Models F & G infer relatively small spread in the [Fe/H] abundance of stars within the solar neighbourhood, with the latter model yielding a substantially small spread. This imposes a constraint on the average velocity of the radial gas mixing for elemental (isotopic) homogenization to be $\leq 1$ km s$^{-1}$ as deduced earlier in the section 4.5. The maximum spread in the cumulative distribution is observed in the case of Models H and J. A secondary maximum in the [Fe/H] value is observed in all the simulations (Fig. 10) corresponding to a value of ~-1.

## 5.    Conclusion

We have presented GCE models for the formation and chemical evolution of the Milky Way galaxy over the galactocentric distances of 2-18 kpc using the revised solar abundances. Compared to the conventional GCE models the present work is based on the Monte Carlo approach of simulations. The chemical evolution of the galaxy is performed by evolving an ensemble of numerous generations of stars that gradually contribute to the isotopic inventories of the galaxy. The adopted approach is much simpler than approach based on solving the integro-differential equations for GCE. We have explored distinct accretional scenarios of the galaxy to understand their influence on the GCE. The galaxy is accreted either by a two-infall or three-infall accretional scenario in order to understand the origin of its three components, viz., the halo, the thick and thin discs. The radial and temporal dependences of the star formation efficiencies are studied to predict the star formation history of the galaxy. The temporal evolution of the elemental abundance gradients for the various elements are compared with the observed elemental abundance distribution of evolved stars. The dynamical processes related with radial gas inflow and radial gas mixing along the galactic disc are studied to understand the influence of these processes on the chemical evolution of the galaxy. The main conclusions drawn from this work are presented in the following.

1.  The galaxy accreted intergalactic material very rapidly during the initial stages (<1 Gyr) to form halo, and later on the disc was formed, with the thin disc forming over much longer timescales with the inside-out accretional scenario. The three-infall accretional model seems to be most viable scenario that explains the majority of the GCE features related with the elemental abundance distributions and gradients. The scenario specifically explains the elemental abundance distribution of the thick disc stars. In its initial stages, the galaxy underwent a high star formation rate which led to the rapid elemental enrichment of the inner regions of the galaxy. The [X/Fe] vs. [Fe/H] evolution predicts the timescale for halo-thick and thin disc phases.

2. The incorporation of the radial gas mixing leads to redistribution of the nucleosynthetic inventories among the adjacent galactic annular rings. This leads to the reduction in the steepness of the elemental abundance gradients and the inferred spread in the metallicity and the [Fe/H] values across the galaxy. We found an agreement with the existing values of the radial gas mixing velocity in the range of 0.1-1 km s$^{-1}$. Based upon our results we rule out the possibilities of >1 km s$^{-1}$ for the average radial mixing velocity of gas.

3. The incorporation of the radial gas inflow across the galaxy with a desired velocity profile reproduces the early observed inversion in the elemental abundance gradients around 2 Gyr.

4. The SN Ia DTD with a 10-15% prompt (<100 Myr) SN Ia population can explain majority of the GCE trends.

5. Finally, it should be pointed out that all the elemental evolutionary trends cannot be explained with the GCE models in a self-consistent manner. There is a substantial scope of further development, specially, in the field of stellar nucleosynthesis.

**Acknowledgement:** We are extremely grateful to the numerous comments and suggestions made by the reviewer that substantially improved the manuscript. The comments made by the reviewer motivated us to perform a wide-range of simulations. This work is supported by PLANEX (ISRO) research grant.

**Table 1.** The simulation models and the associated parameters.

| S.No. | Model | Simulation features | Accretion model | Halo Surface mass density ($M_\odot$ pc$^{-2}$) | DTD* for SN Ia | Figure references§ |
|-------|-------|---------------------|-----------------|------------------------------------------------|----------------|--------------------|
| 1 | Model A | $\alpha$, $\kappa$ | Two-infall | 17 up to 10 kpc & $r^{-1}$ dependence outside | MR01 | Fig. 1, 4, 7, 10 |
| 2 | Model B | $\alpha$, $\kappa$ | Two-infall | Constant 17 | MR01 | (Supp.) Fig. 11, 12, 13 |
| 3 | Model C | $\beta$, $\kappa$ | Two-infall | Same as Model A | MR01 | (Supp.) Fig. 14, 15, 16 |
| 4 | Model D | $\alpha$, $\kappa$ | Two-infall | Same as Model A + 20 %-disc mass accreted in Halo | MR01 | (Supp.) Fig. 17, 18, 19 |
| 5 | Model E | $\alpha$, $\kappa$ | Two-infall | Same as Model A | G05 | (Supp.) Fig. 20, 21, 22 |
| 6 | Model F | $\alpha$, $\lambda$ | Two-infall | Same as Model A | MR01 | (Supp.) Fig. 23, 24, 25 |
| 7 | Model G | $\alpha$, $\mu$ | Two-infall | Same as Model A | MR01 | (Supp.) Fig. 26, 27, 28 |
| 8 | Model H | $\alpha$, $\kappa$ | Three-infall | Same as Model A | MR01 | Fig. 2, 5, 8, 10 |
| 9 | Model I | $\beta$, $\delta$, $\kappa$ | Two-infall | 17 up to 10 kpc & 1.7 outside | MR01 | (Supp.) Fig. 29, 30, 31 |
| 10 | Model J | $\beta$, $\delta$, $\lambda$ | Two-infall | Same as Model I | MR01 | Fig. 3, 6, 9, 10 |

$\alpha$: Radial dependence of star formation efficiency, $\eta(r)$, with a value of 4, 3, 2, 1, 0.7, 0.45, 0.3 and 0.2 from the 1$^{st}$ to the 8$^{th}$ annular rings.

$\beta$: Assumed value of unity for the radial stellar formation efficiency, $\eta(r)$, across the galaxy.

$\kappa$: No radial gas mixing among the adjacent rings.

$\lambda$: Assumed 0.06 % radial gas mixing (with velocity ~1 km s$^{-1}$) among the adjacent rings.

$\mu$: Assumed 0.23 % radial gas mixing (with velocity ~4 km s$^{-1}$) among the adjacent rings.

$\delta$: Radial gas inflow velocities; 0, 0, 0.5, 1, 2, 3 and 3.7 km s$^{-1}$ across the 1$^{st}$ to 7$^{th}$ the rings.

* DTD for supernova SN Ia; Matteucci & Recchi (2001) [MR01] & Greggio (2005) [G05].

§ Figures 11-31 are provided only in the supplementary data.

**Figure Captions:**

**Fig. 1.** The numerically deduced temporal evolution in case of ***Model A*** of, **a)** the star formation rate ($M_\odot$ pc$^{-2}$ Myr$^{-1}$), **b)** [Fe/H], **c)** the metallicity, '$Z$', **d)** the stellar birth rate (in Myr$^{-1}$), **e)** the total surface mass density (stars+interstellar gas), $\sigma_{Total}$ ($M_\odot$ pc$^{-1}$), **f)** the surface mass density of the various *live* stars $\sigma_{stars}$ ($M_\odot$ pc$^{-1}$), **g)** the propagation of the modulus of the cumulative errors, $|\{(\sigma_{Total} - \sigma_{gas}) - (\sigma_{stars} + \sigma_{White\,dwarf}\ \sigma_{Neutron\,star + Black\,hole})\}$ Ring area| (in $M_\odot$ pc$^{-1}$), **h)** the remnant surface mass density of Neutron stars + Black holes, **i)** the surface mass density of white dwarfs, are presented for the various annular rings. **j-l)** The deduced temporal evolution of the supernovae (SN Ia, SN II, and Ib/c) rates for the various annular rings are also presented. The modulus of the cumulative error for the galaxy (2-18 kpc) is < 0.15 $M_\odot$.

**Fig. 2.** Identical to Fig. 1 for Model H.

**Fig. 3.** The numerically deduced temporal evolution in case of ***Model J*** of, **a)** the star formation rate ($M_\odot$ pc$^{-2}$ Myr$^{-1}$), **b)** [Fe/H], **c)** the metallicity, '$Z$', **d)** the stellar birth rate (in Myr$^{-1}$), **e)** the total surface mass density (stars+interstellar gas), $\sigma_{Total}$ ($M_\odot$ pc$^{-1}$), **f)** the surface mass density of the various *live* stars $\sigma_{stars}$ ($M_\odot$ pc$^{-1}$), **g)** the gas surface mass density, $\sigma_{Gas}$ ($M_\odot$ pc$^{-1}$), **h)** the remnant surface mass density of Neutron stars + Black holes, **i)** the surface mass density of white dwarfs, are presented for the various annular rings. **j-l)** The deduced temporal evolution of the supernovae (SN Ia SN II, and Ib/c) rates for the various annular rings are also presented.

**Fig. 4.** The normalized elemental abundance gradients of some selective elements across the galaxy at seven distinct epochs of the evolution in the case of ***Model A***. The observational data is based on Cepheid variables (C), Main sequence Stars (MS), Dwarf stars (D), Giants (G), Subgiants (SB), Red giants (RG), Sub Red-Giants (SRG) PNe, H II region, Open Cluster (OC), Globular Cluster (GC), Sub-Dwarf (SD) and OB stars. The references for the various symbols used for data along with the figure are, Green Stars (Simpson et al. 1995) (H II Region) & (Afflerbach et al. 1997) (H II Region); Brown plus (Twarog et al. 1997) (OC) & (Carraro et al. 1998) (F & G stars) & (Chen et al. 2003) (OC); Blue filled inverted triangles (Gummersbach et al. 1998) (B stars); Red filled triangle (Daflon & Cunha 2004) (C); Cyan filled circle (Andrievsky et al. 2002b, 2004) (C); Blue diamonds (Yong et al. 2006) (C); Blue open circle (Cescutti et al. 2007) (C); Green ellipse (Boeche et al. 2013) (D); Pink filled circle (Lemasle et al. 2013) (C); Grey filled circle (Martin et al. 2015) (C); Yellow filled circle (Cunha et al. 2016) (RG). The normalization in all cases is done with respect to the solar abundance at 9 Gyr within the solar annular ring.

**Fig. 5.** Identical to Fig. 4 for Model H.

**Fig. 6.** Identical to Fig. 4 for Model J.

**Fig. 7.** The temporal evolution of the normalized [X/Fe] vs. [Fe/H] in the case of ***Model A***. The normalization in all cases is done with respect to the solar abundance at 9 Gyr within the solar annular ring. The observational data is also presented for comparison (Clegg et al. 1981 (MS); Gratton 1985 (G); Laird 1985 (F & G dwarf); Luck and Bond 1985 (G); Gratton & Ortolani 1986 (Late-type star); Tomkin et al. 1986 (D); Carbon et al. 1987 (D); François 1987, 1988

(D); Gratton & Sneden 1988, 1991 (G); Hartmann & Gehren 1988 (SD); Sneden & Crocker 1988 (D &SG); Barbuy & Erdelyi-Mendes 1989 (G); Magain 1989 (D); Peterson et al. 1990 (G); Zhao & Magain 1990 (D); Meusinger et al. 1991 (MS & D); Sneden et al. 1991 (GC); Edvardsson et al. 1993 (D); Nissen et al. 1994 (metal-poor stars); McWilliam et al. 1995 (metal-poor stars); Ryan et al. 1996 (D & G); Israelian et al. 1998 (metal poor stars); Boesgaard et al. 1999 (metal-poor stars); Stephens 1999 (D); Carretta et al. 2000 (D); Chen et al. 2000 (D); Mishenina et al. 2000 (D); Rocha-Pinto et al. 2000 (D); Goswami & Prantzos 2000; Gratton et al. 2003 (SD), Reddy et al. 2003 (D); Cayrel et al. 2004 (D&G); Mishenina 2004 (D); Soubiran & Girard 2005 (FGK stars); Nomoto et al 2006; Reddy et al. 2006 (D); Shi et al. 2009 (D); Ruchti et al. 2011 (RG, MS & SRG); Adibekyan et al. 2012 (FGK Dwarf); Bensby et al. 2014 (F & G Dwarf and SG). The abbreviation for the various stellar types is identical to the one used in Fig. 4. The broad regions associated with the halo stars, thick and thin disc stars are marked on the figures.

**Fig. 8.** Identical to Fig. 7 for Model H.

**Fig. 9.** Identical to Fig. 7 for Model J.

**Fig. 10. (a-f)** The G-dwarf distribution of [Fe/H] for the 0.8-1.25 $M_\odot$ stars at the present epoch corresponding to ~13.5 Gyr for all the annular rings for the Models A, D, F, G, H and J. The observational data in the solar neighborhood is taken from Holmberg et al. 2007 (green histogram) and Jørgensen 2000 (brown histogram) for comparison.

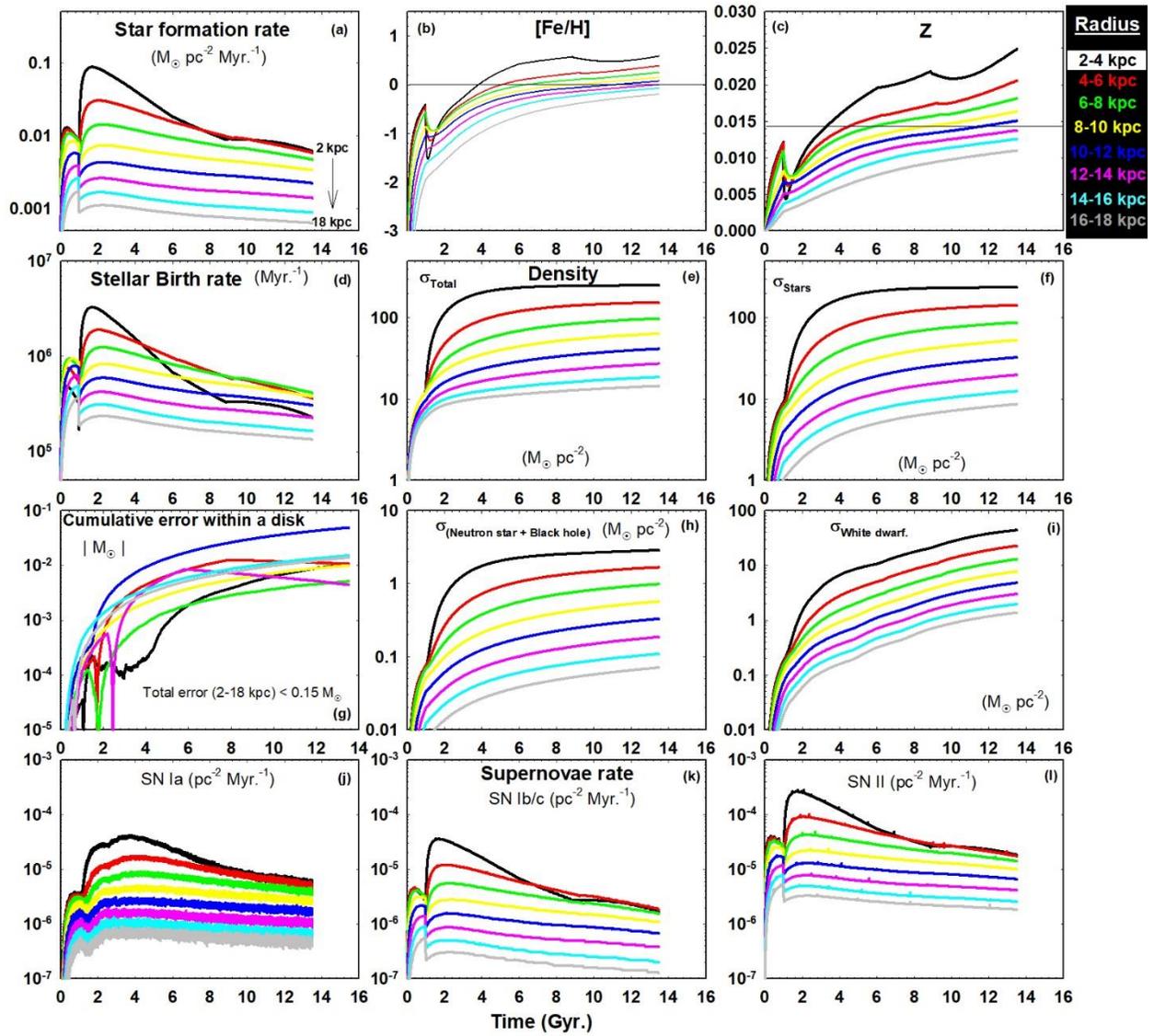

**Fig. 1**

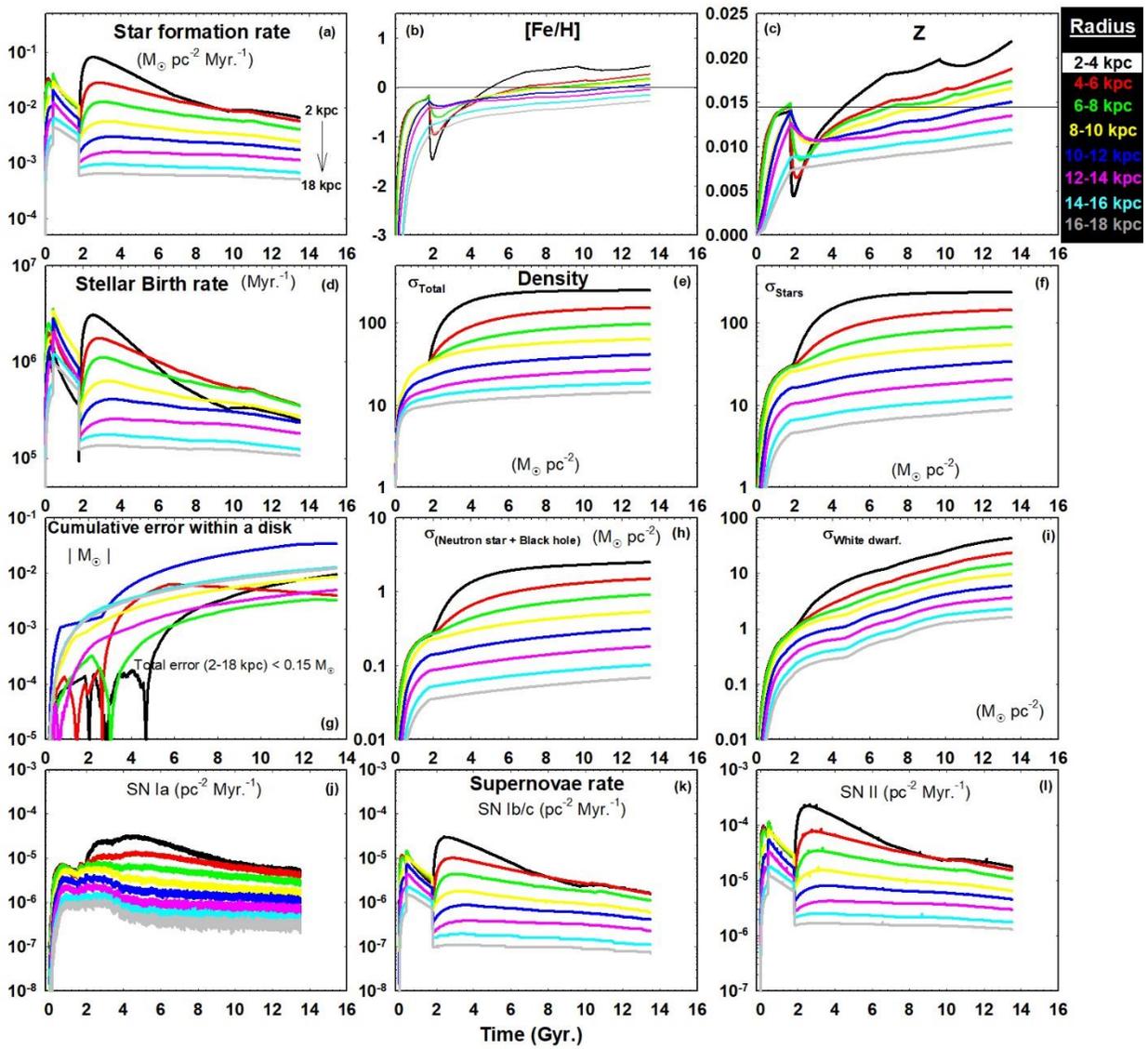

**Fig. 2**

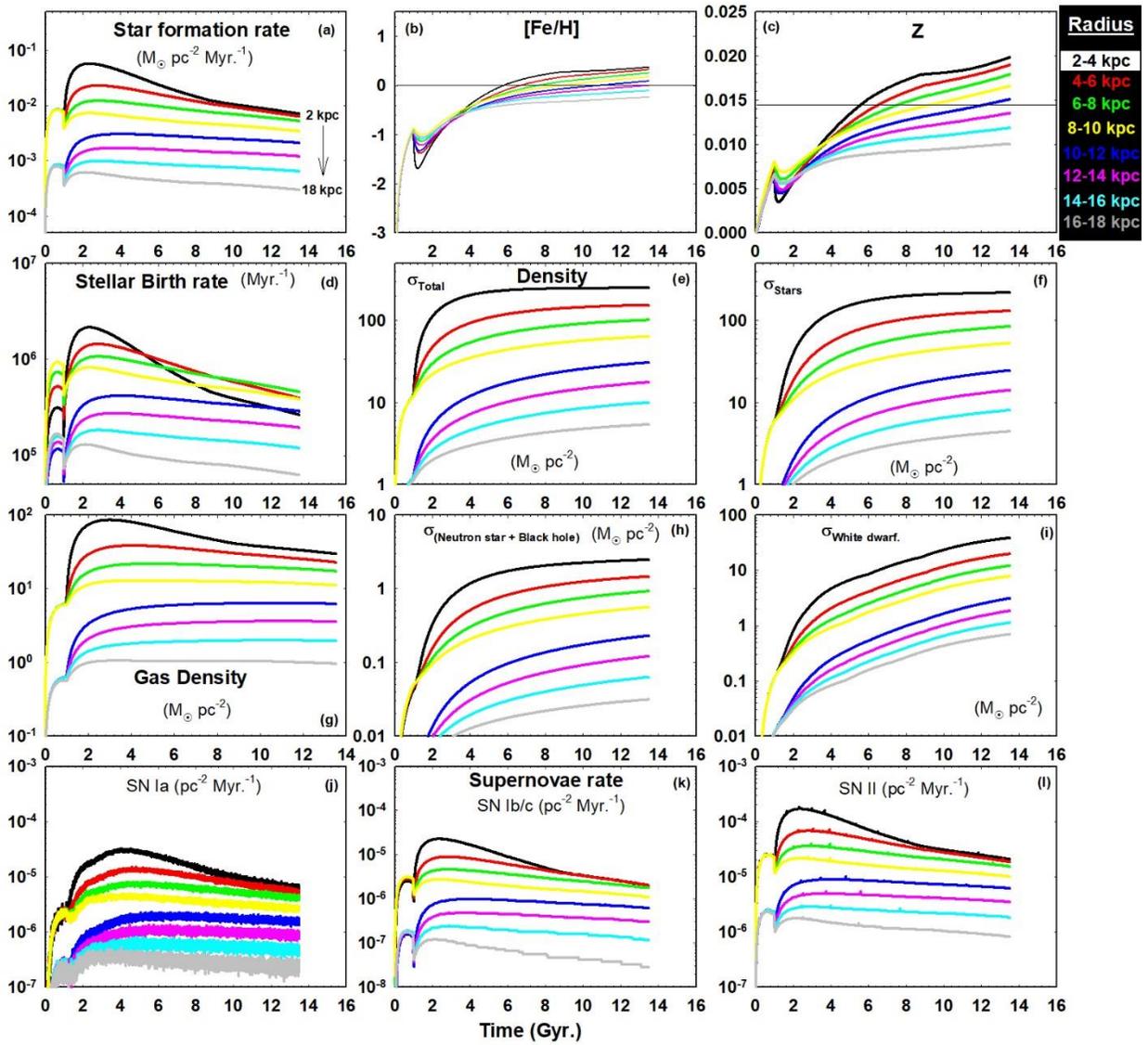

**Fig. 3**

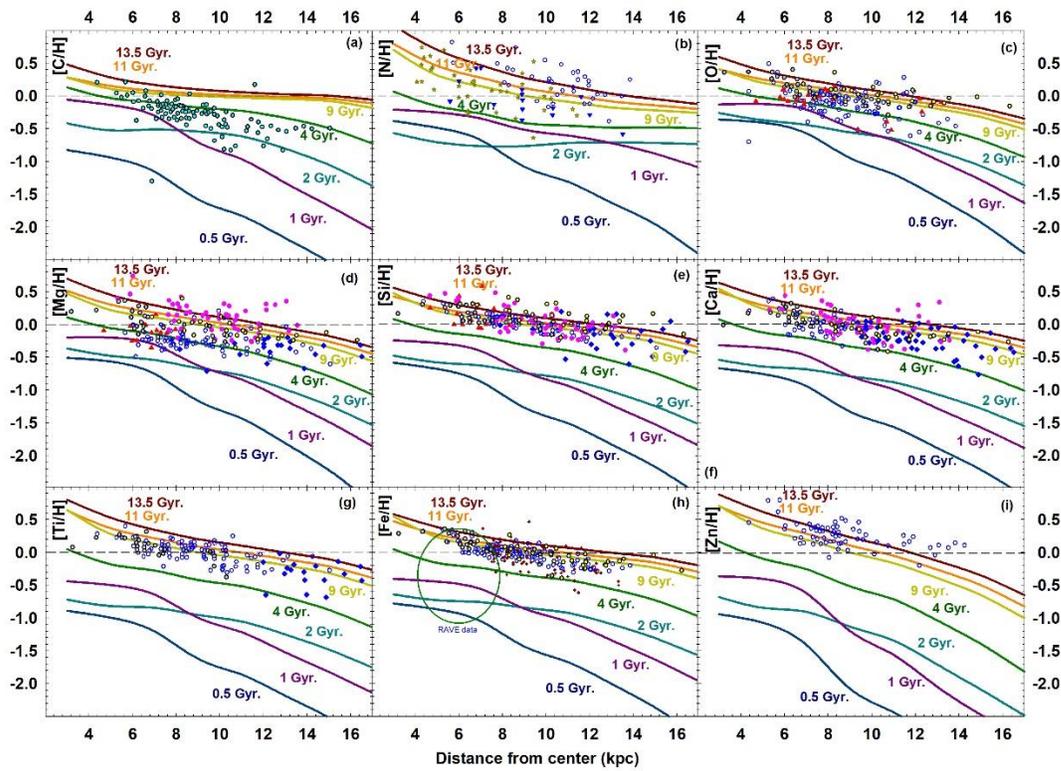

**Fig. 4**

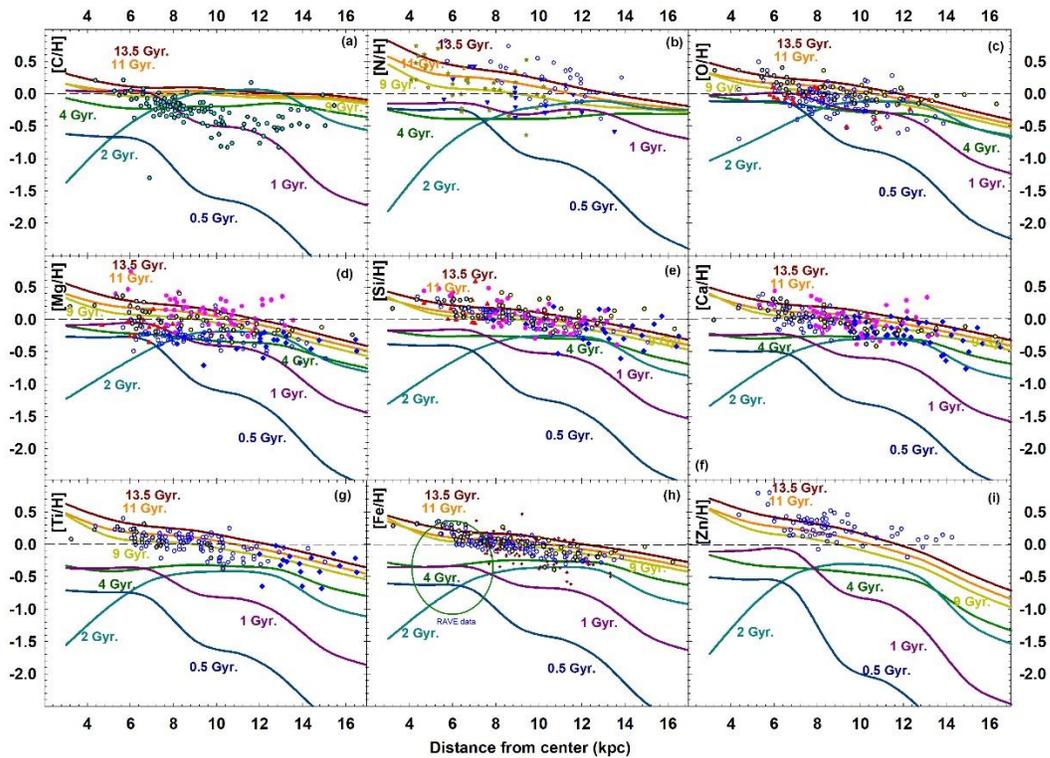

**Fig. 5**

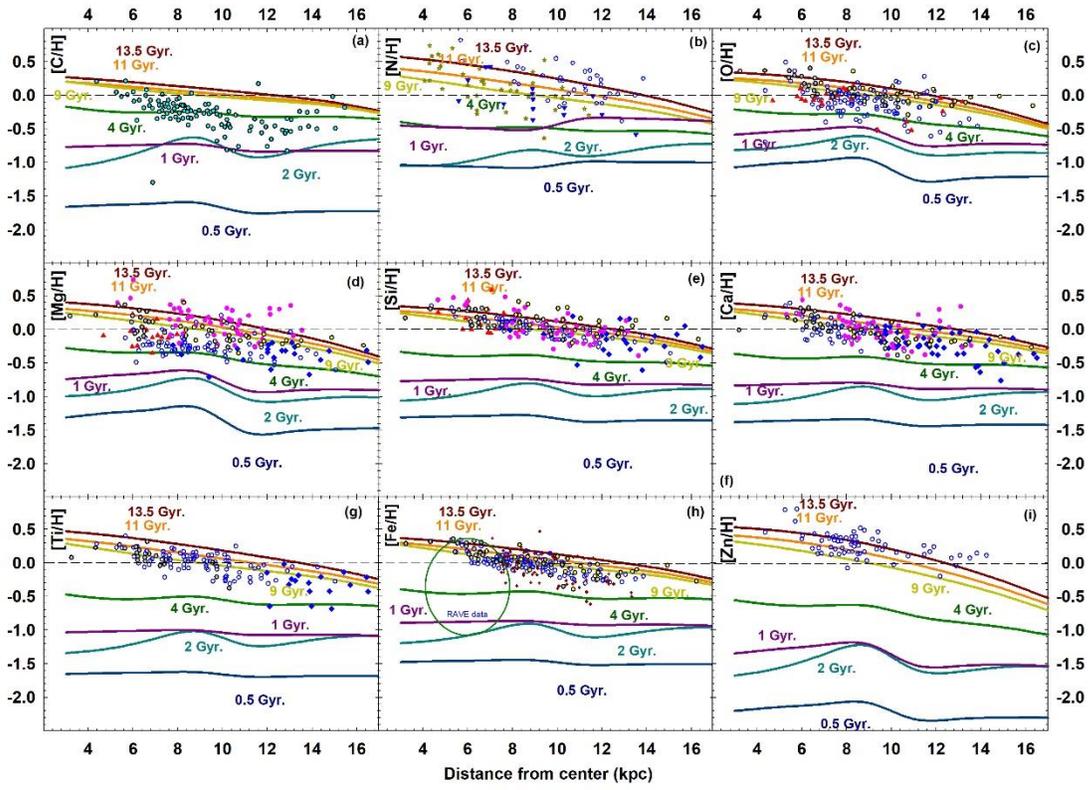

**Fig. 6**

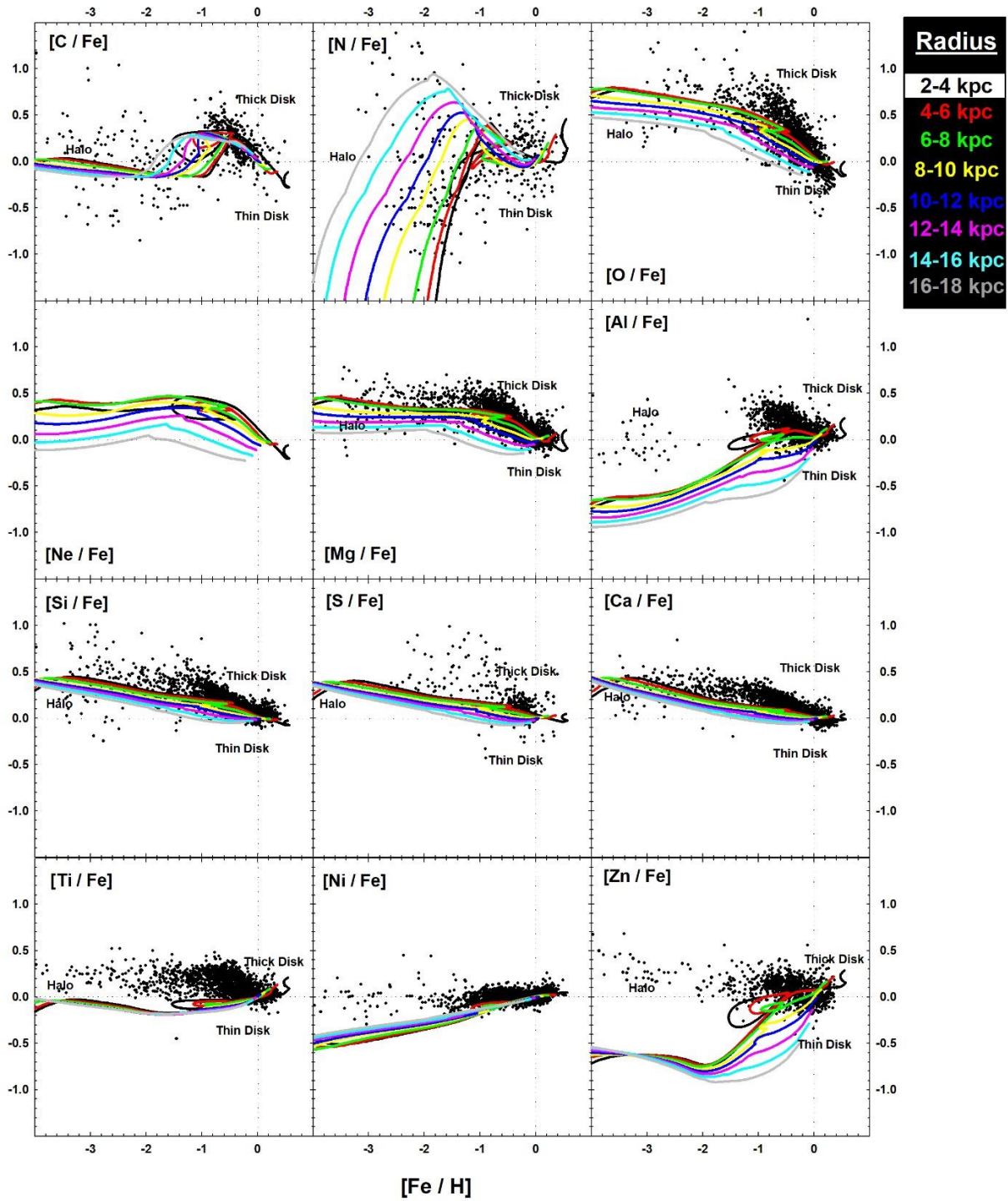

**Fig. 7**

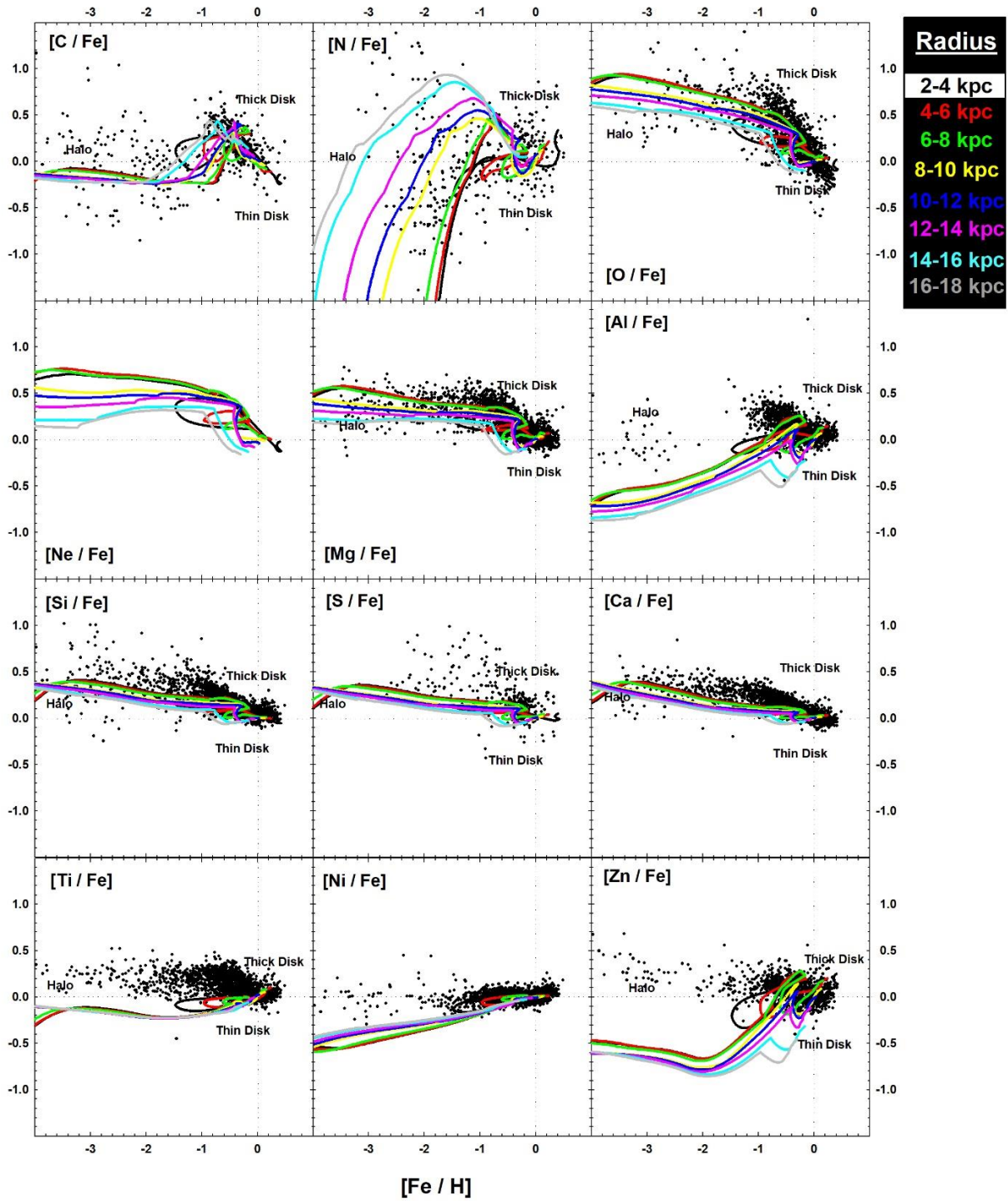

**Fig. 8**

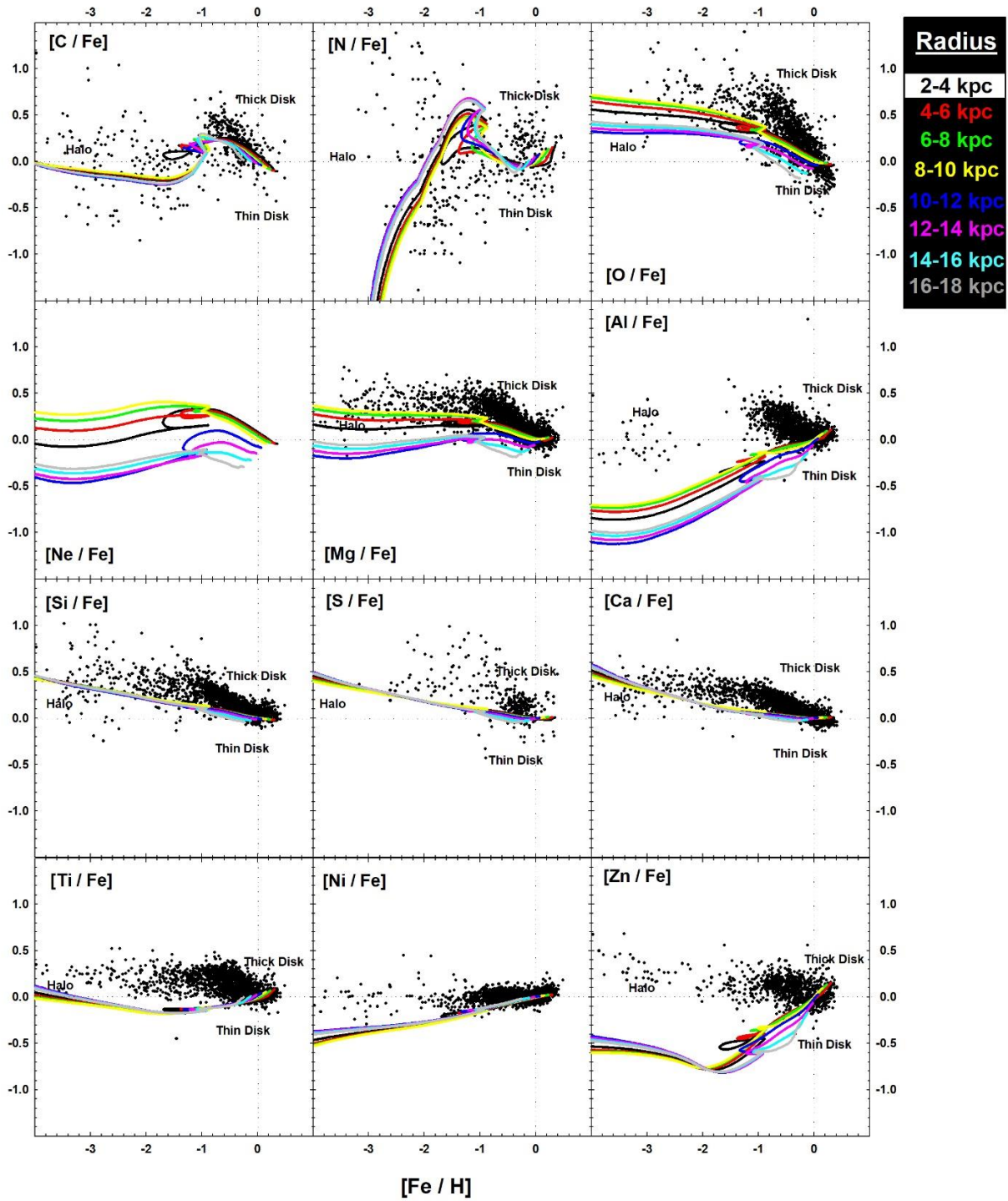

**Fig. 9**

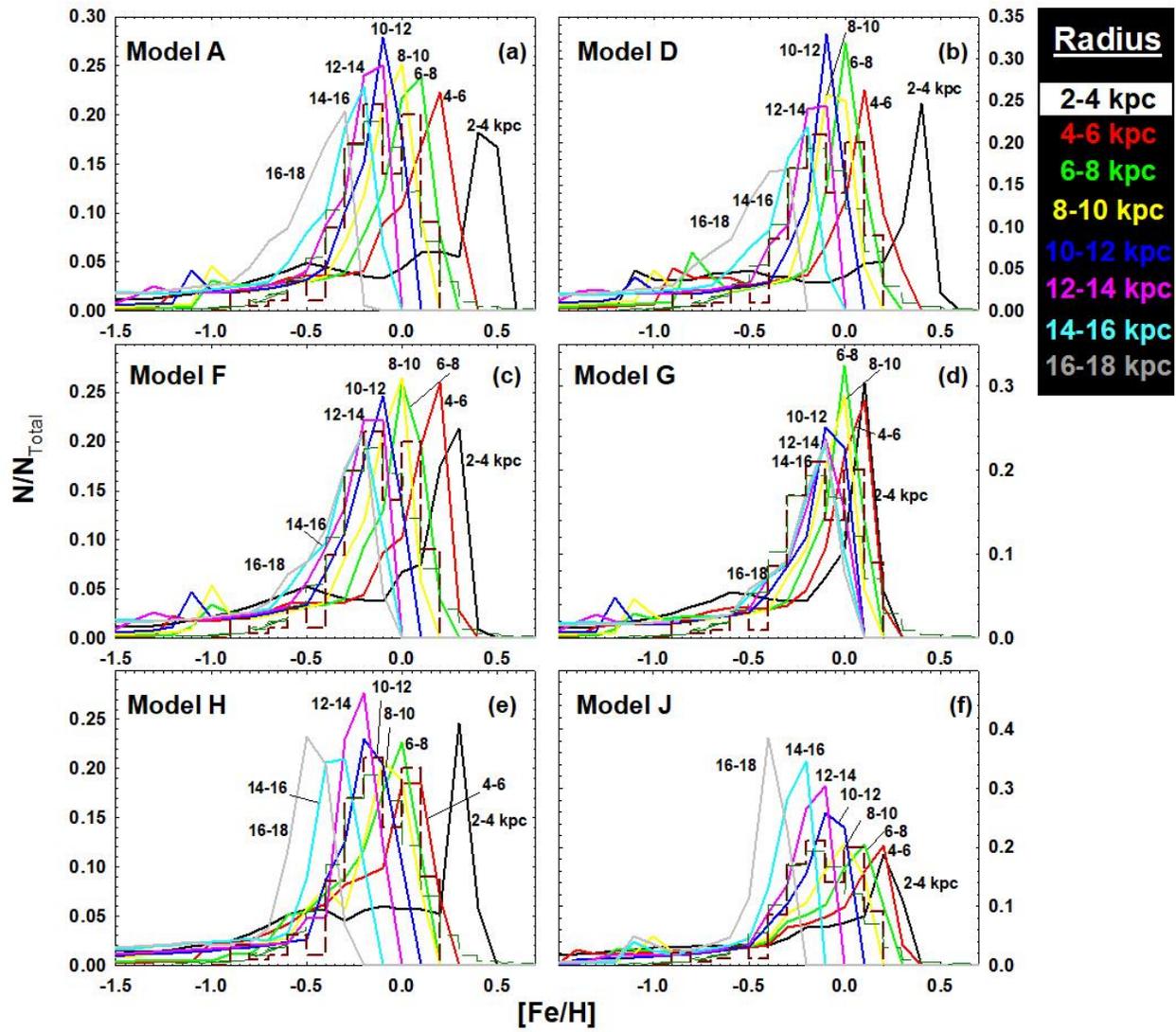

**Fig. 10**

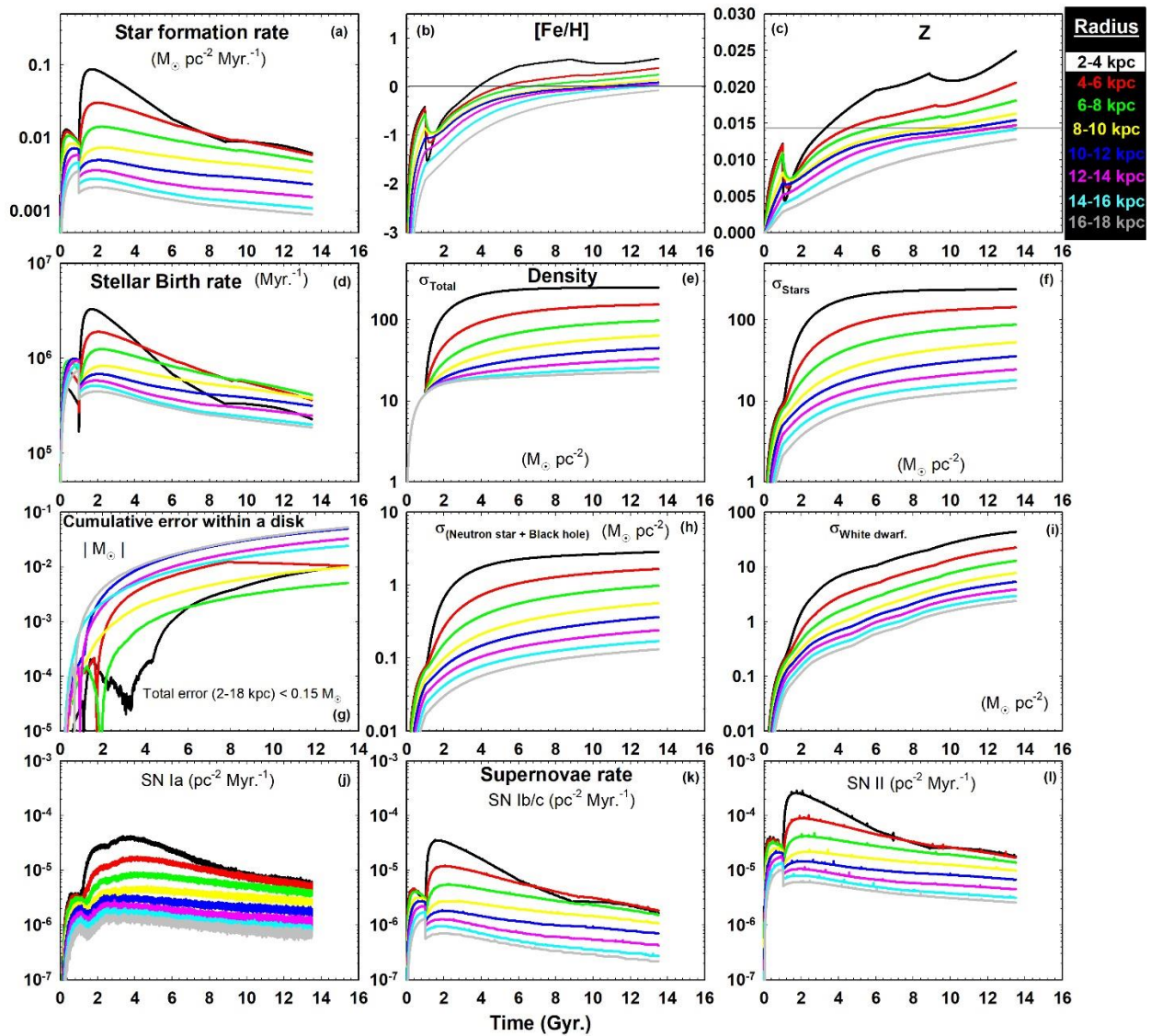

**Fig. 11.** The numerically deduced temporal evolution in case of ***Model B*** of, **a)** the star formation rate (M$_\odot$ pc$^{-2}$ Myr$^{-1}$), **b)** [Fe/H], **c)** the metallicity, '$Z$', **d)** the stellar birth rate (in Myr$^{-1}$), **e)** the total surface mass density (stars+interstellar gas), $\sigma_{Total}$ (M$_\odot$ pc$^{-1}$), **f)** the surface mass density of the various *live* stars $\sigma_{stars}$ (M$_\odot$ pc$^{-1}$), **g)** the propagation of the modulus of the cumulative errors, $|\{(\sigma_{Total} - \sigma_{gas}) - (\sigma_{stars} + \sigma_{White\,dwarf}\ \sigma_{Neutron\,star + Black\,hole})\}$ Ring area| (in M$_\odot$ pc$^{-1}$), **h)** the remnant surface mass density of Neutron stars + Black holes, **i)** the surface mass density of white dwarfs, are presented for the various annular rings. **j-l)** The deduced temporal evolution of the supernovae (SN Ia, SN II, and Ib/c) rates for the various annular rings are also presented. The modulus of the cumulative error for the galaxy (2-18 kpc) is < 0.15 M$_\odot$. See Table 1 for details.

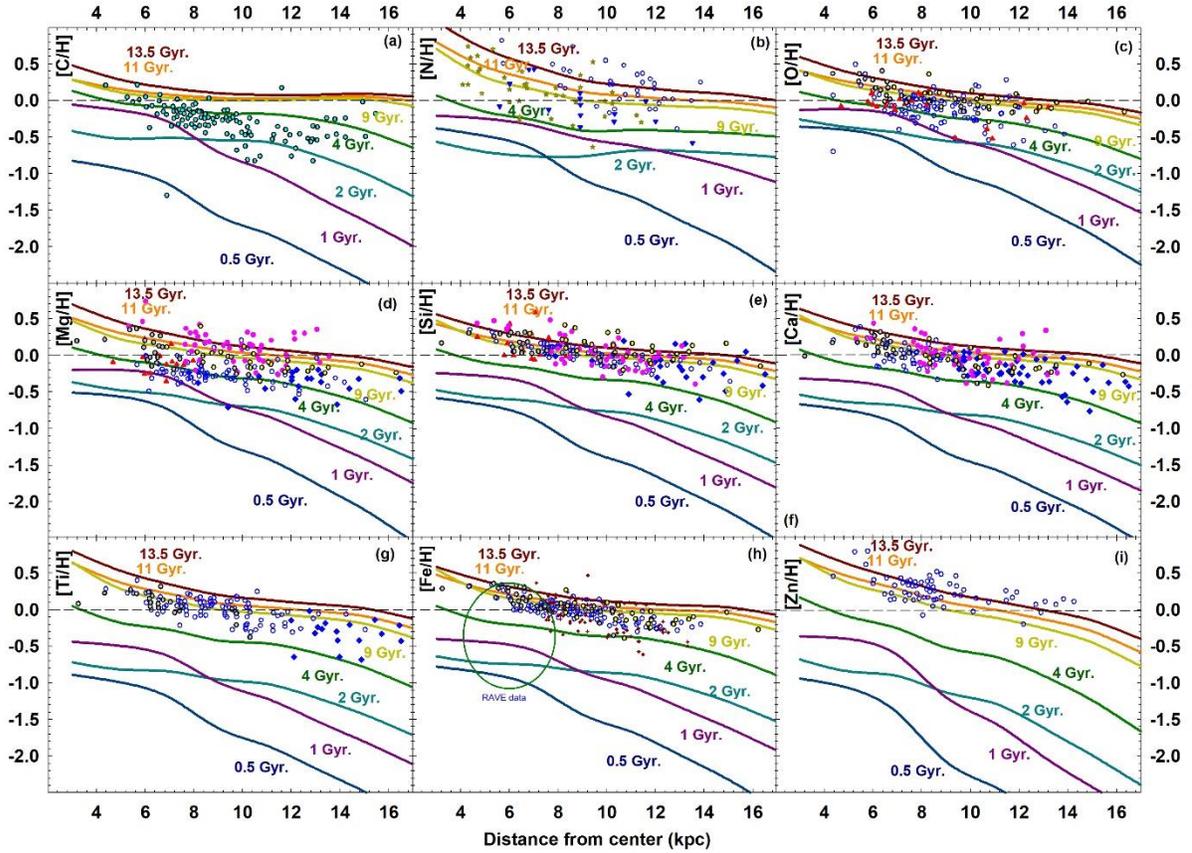

**Fig. 12.** The normalized elemental abundance gradients of some selective elements across the galaxy at seven distinct epochs of the evolution in the case of ***Model B***. The observational data is identical to the one presented in Fig. 4. See Table 1 for details.

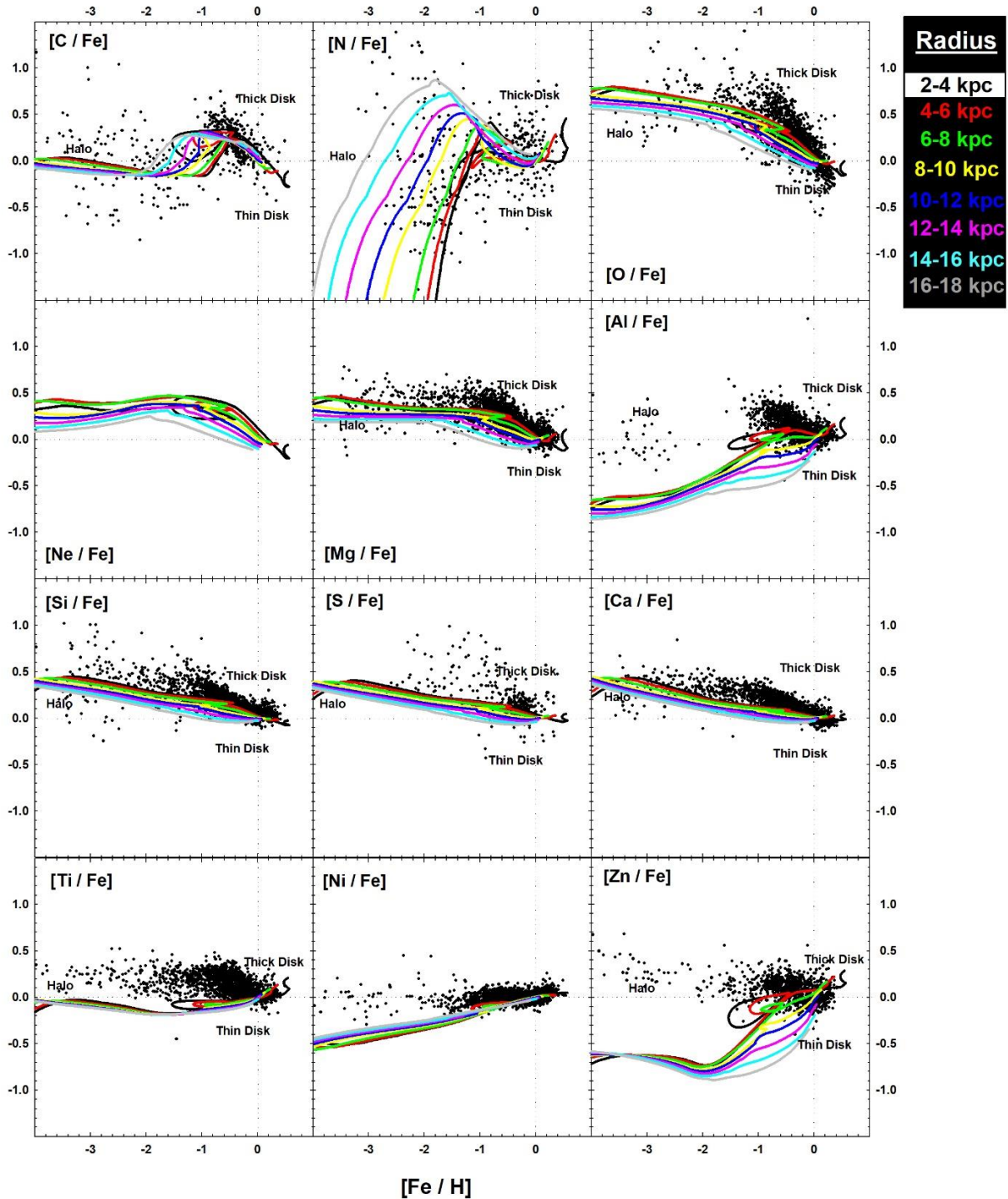

**Fig. 13.** The temporal evolution of the normalized [X/Fe] vs. [Fe/H] in the case of **_Model B_**. The normalization in all cases is done with respect to the solar abundance at 9 Gyr. within the solar annular ring. The observational data is identical to one presented in Fig. 7. See Table 1 for details.

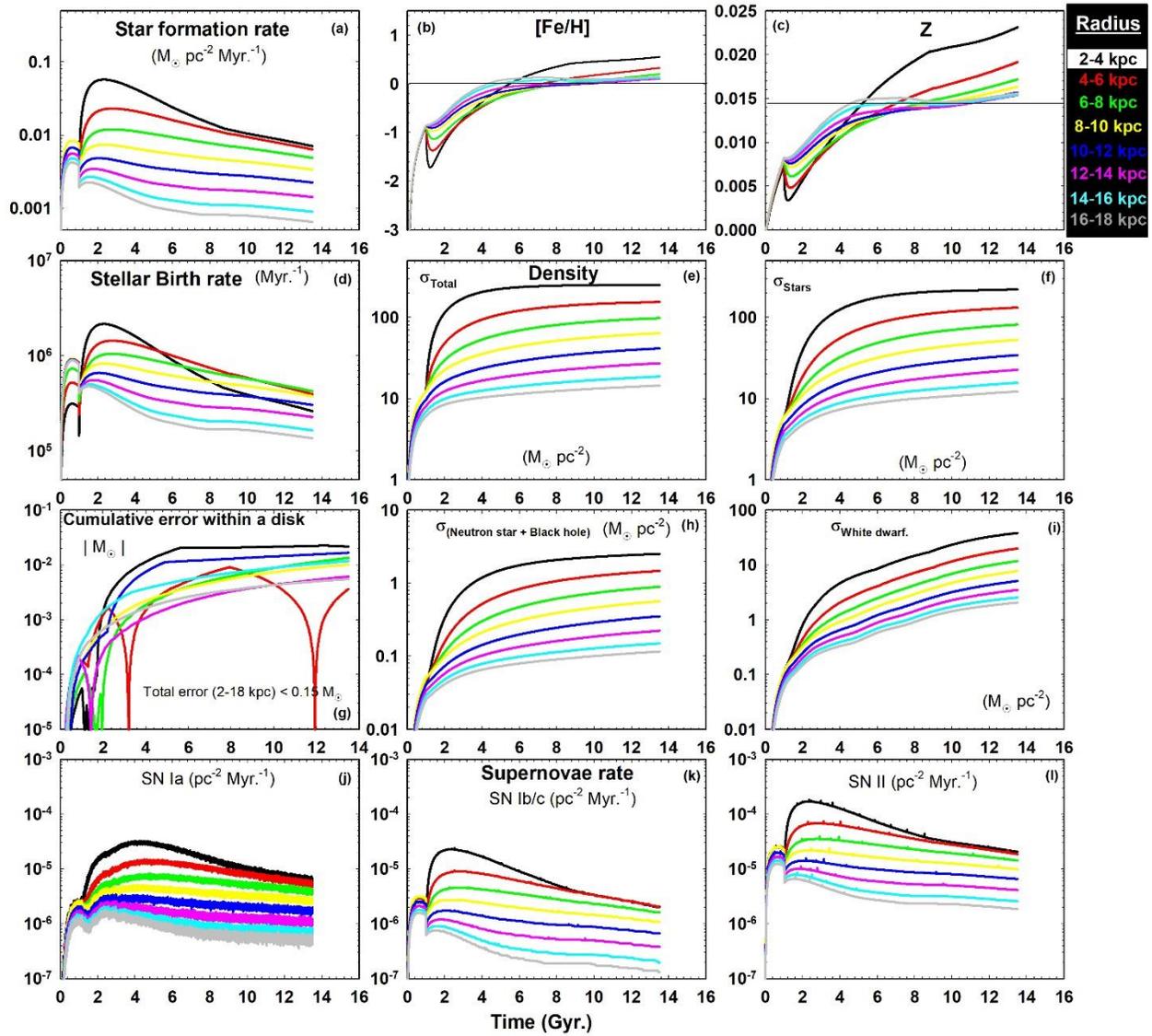

**Fig. 14.** Identical to Fig. 11 for **Model C**.

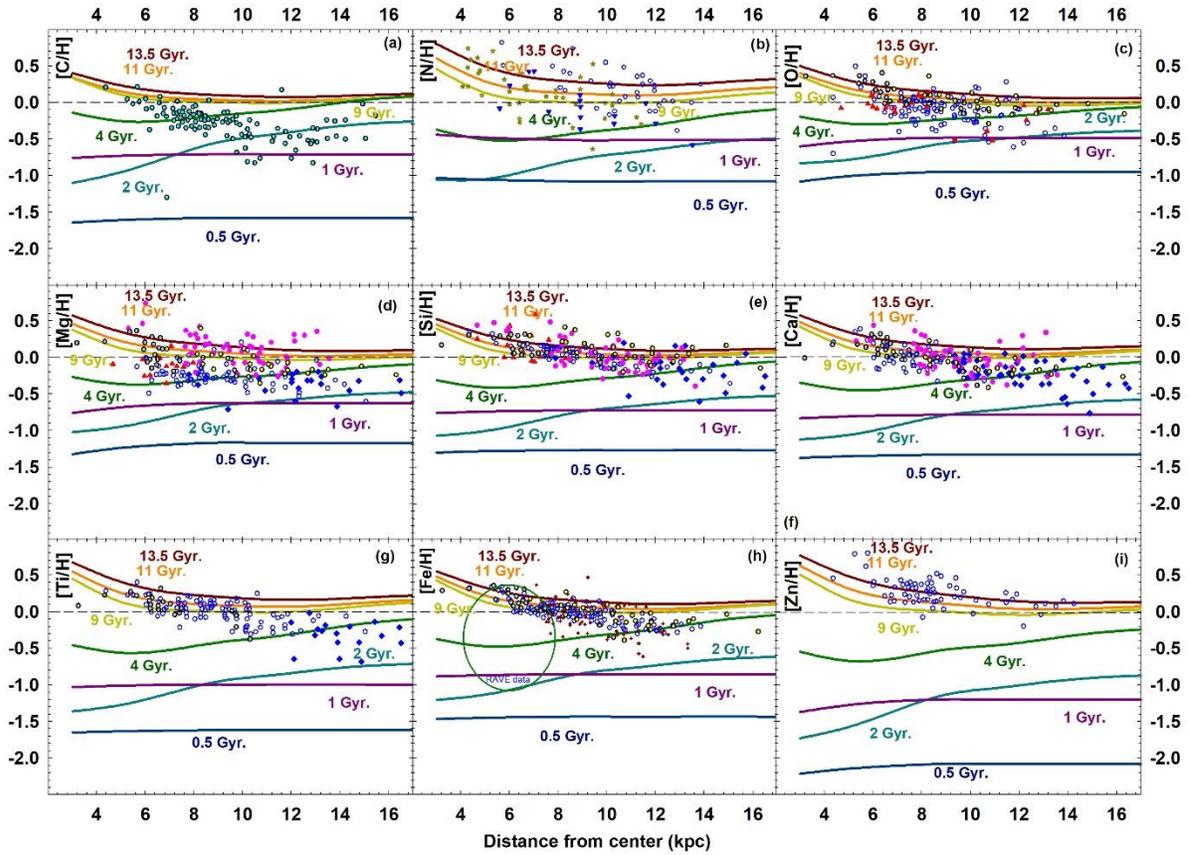

**Fig. 15.** Identical to Fig. 12 for **_Model C_**.

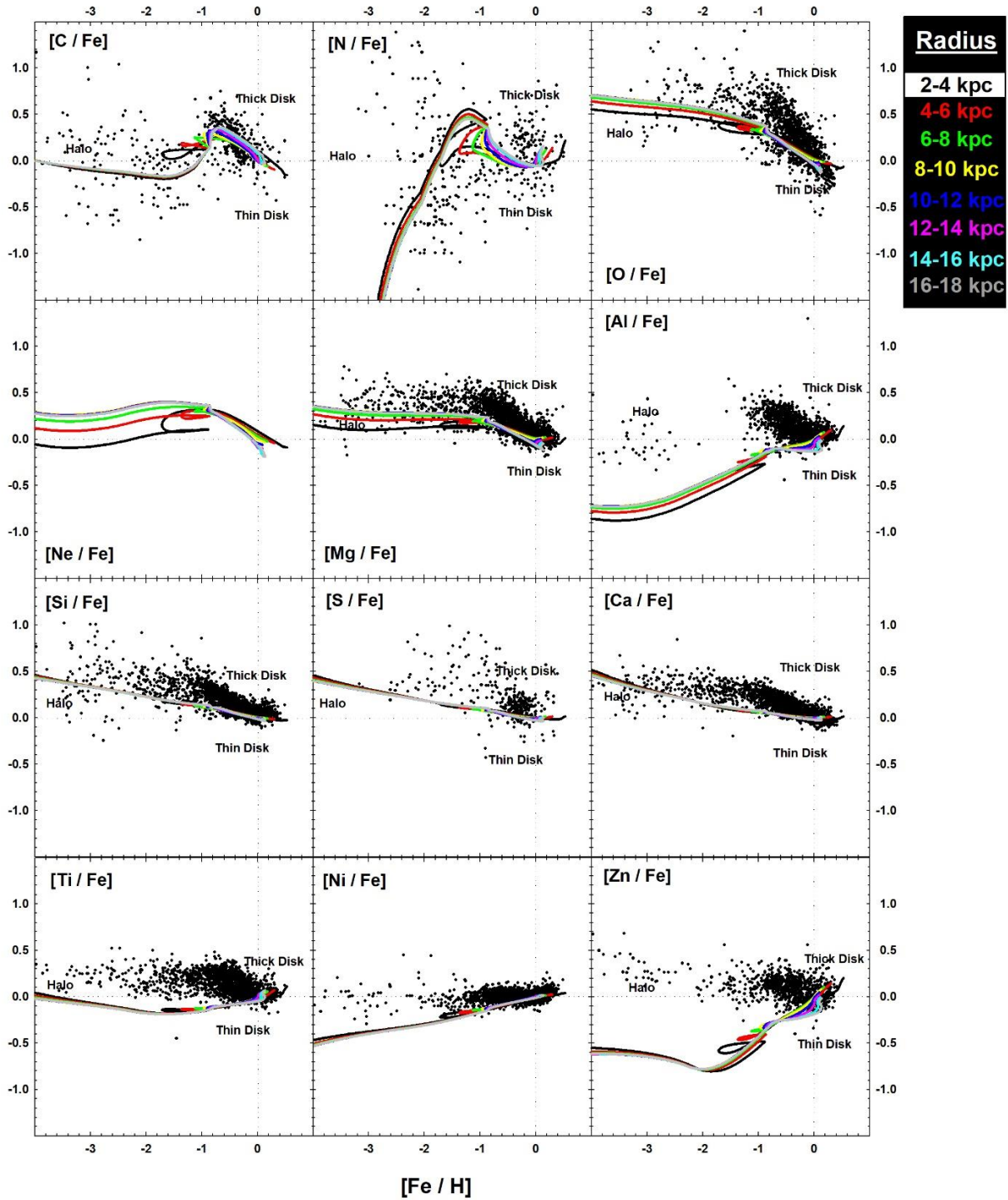

**Fig. 16.** Identical to Fig. 13 for *Model C*.

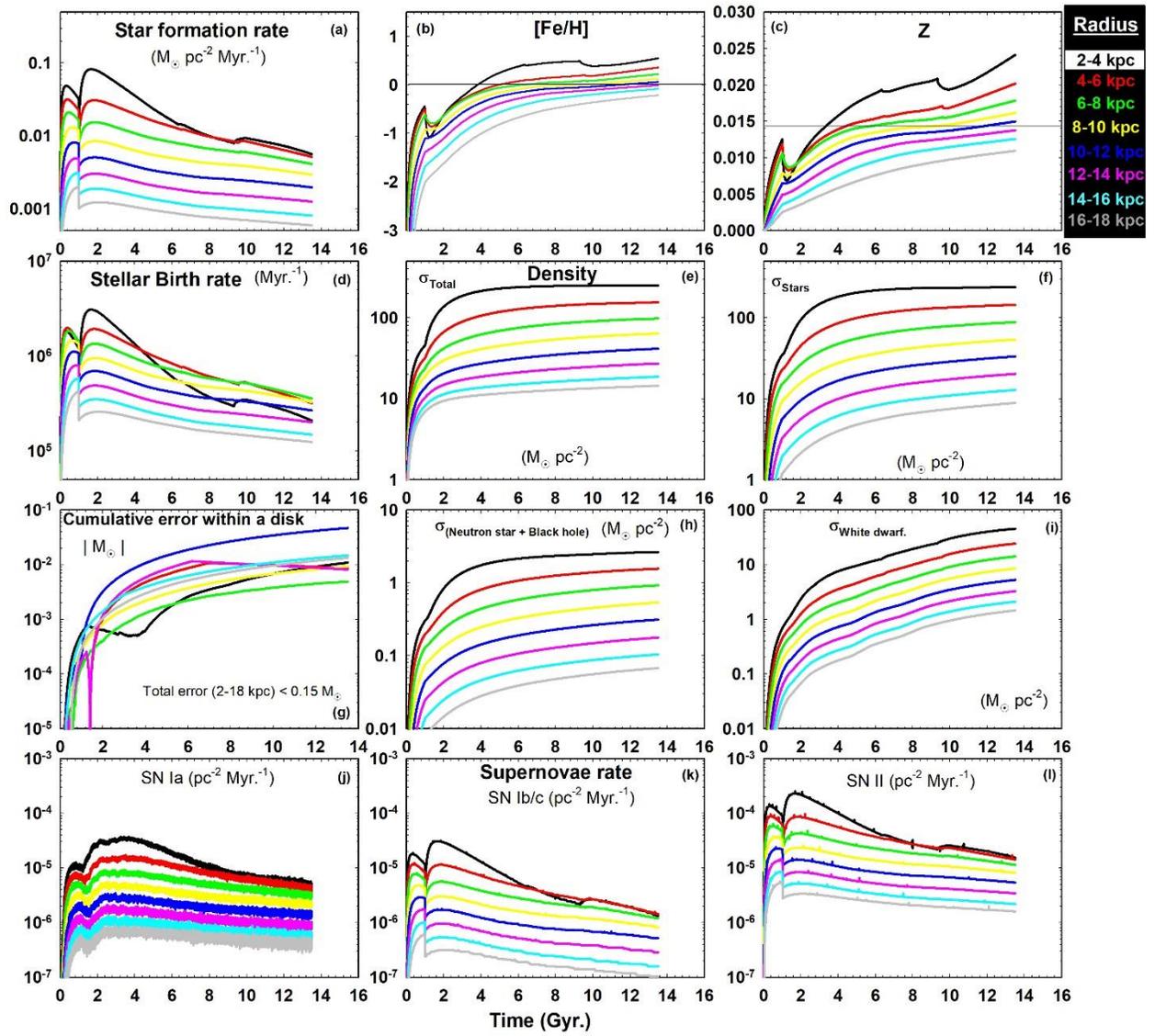

**Fig. 17.** Identical to Fig. 11 for *Model D*.

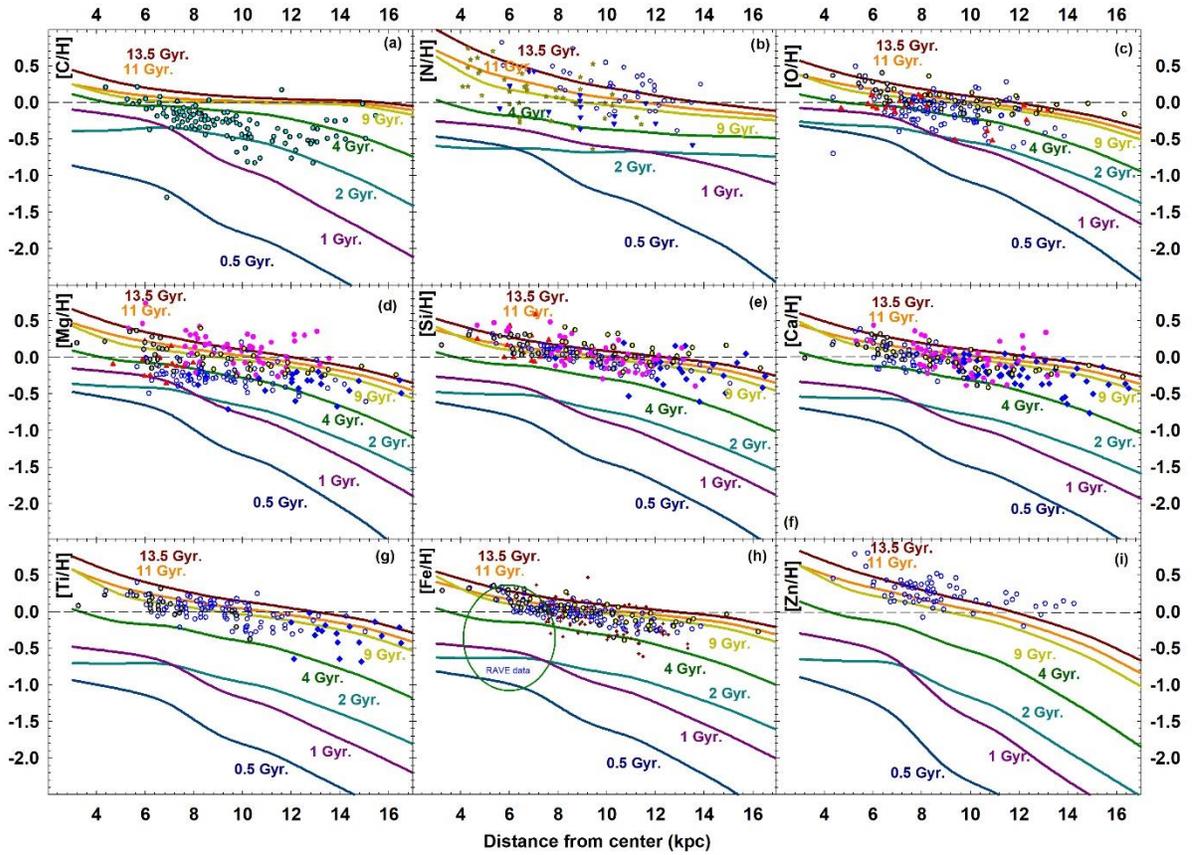

**Fig. 18.** Identical to Fig. 12 for **_Model D_**.

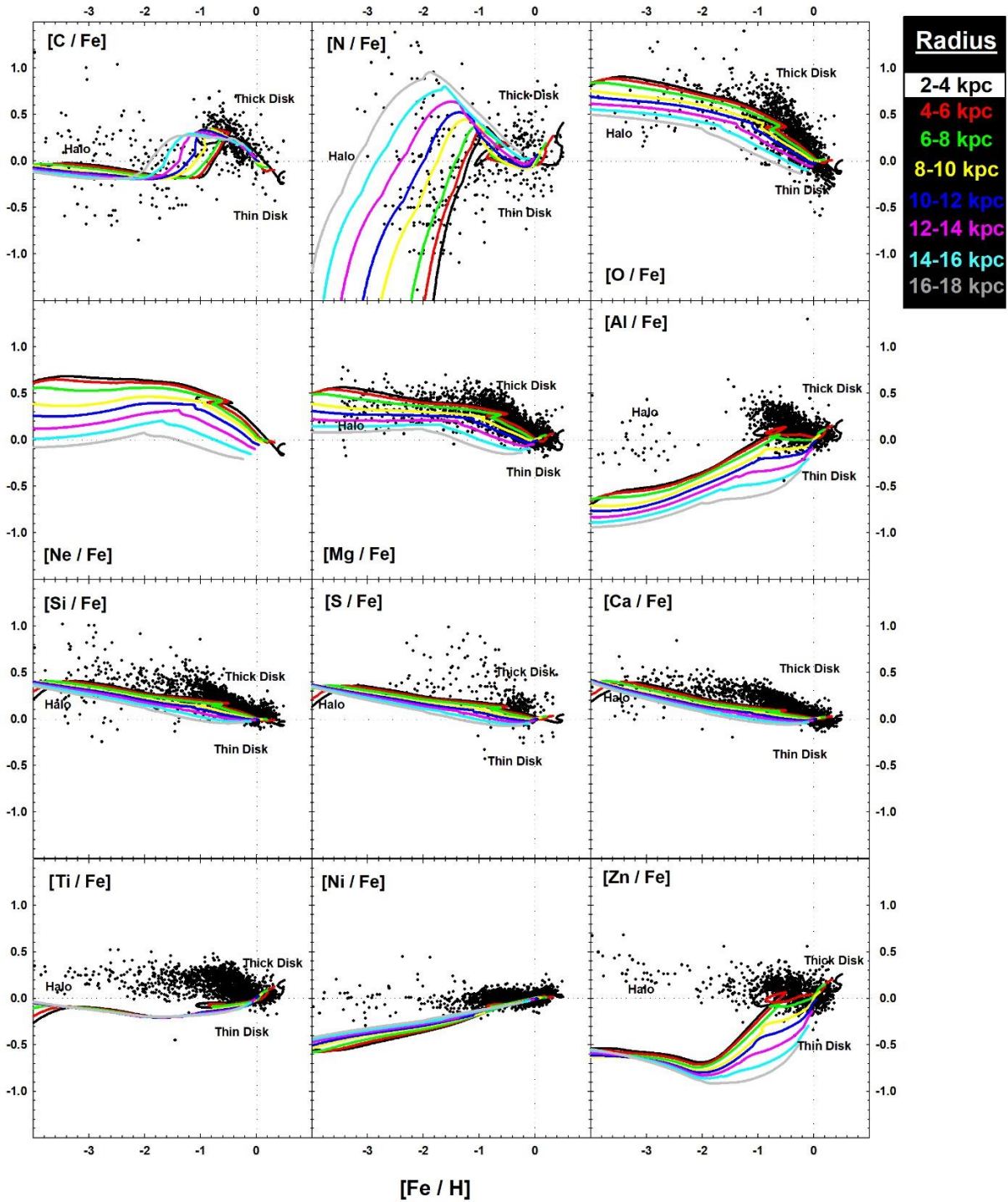

**Fig. 19.** Identical to Fig. 13 for *Model D*.

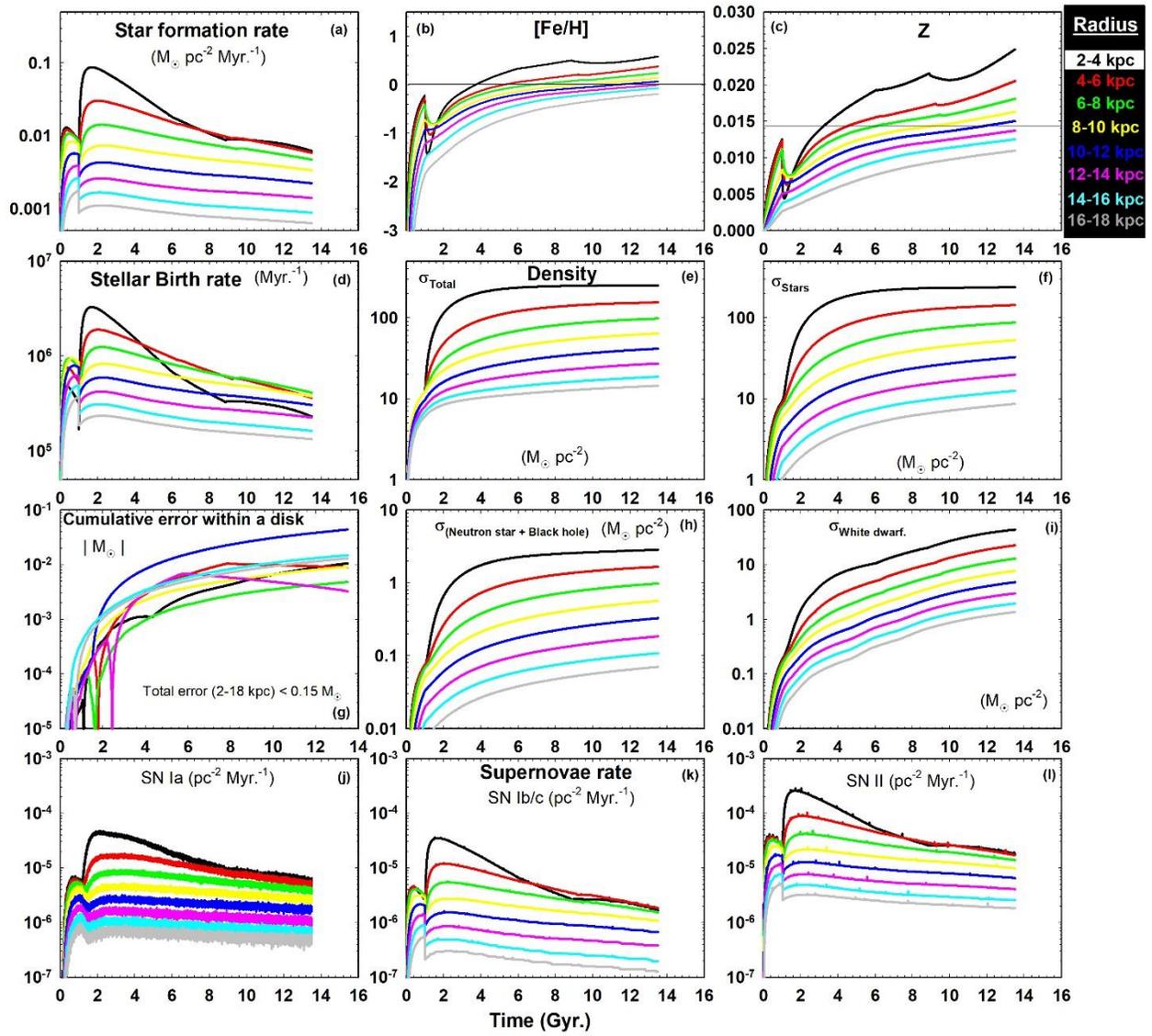

**Fig. 20.** Identical to Fig. 11 for **Model E**.

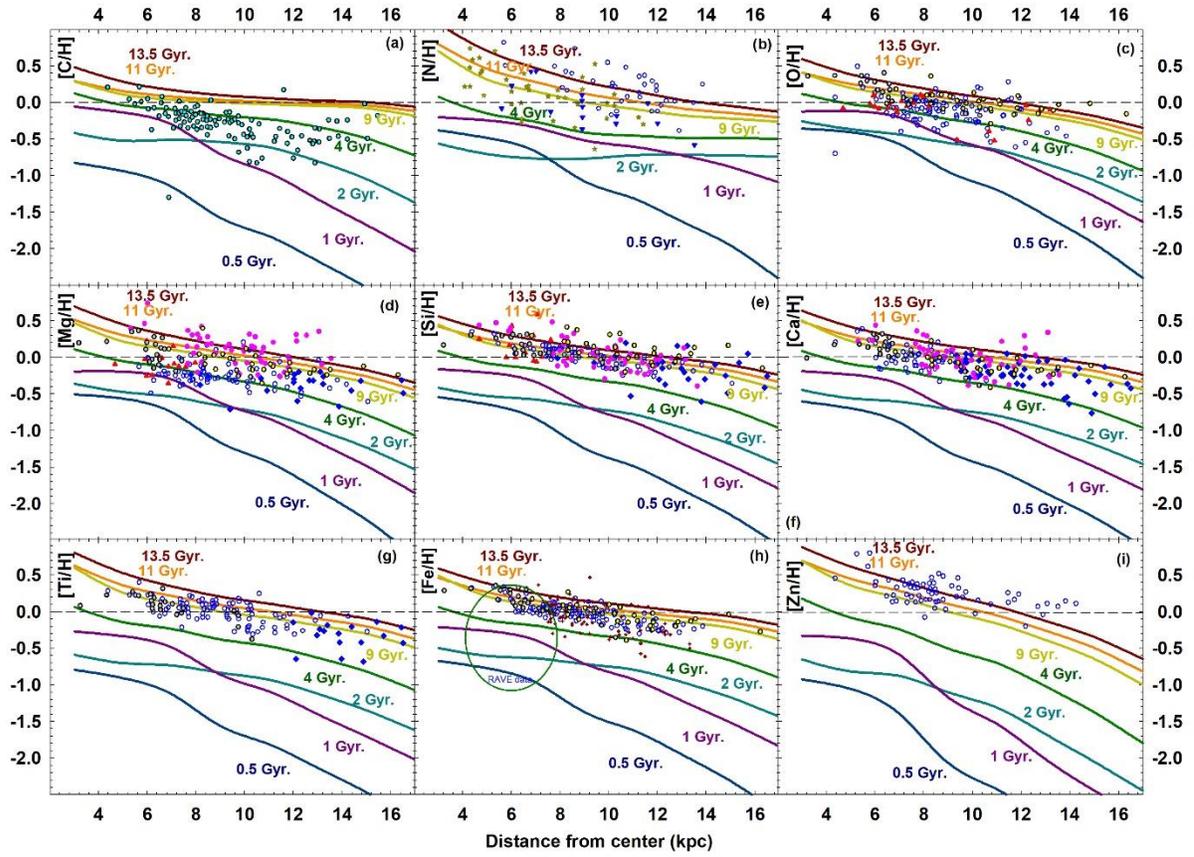

**Fig. 21.** Identical to Fig. 12 for *Model E*.

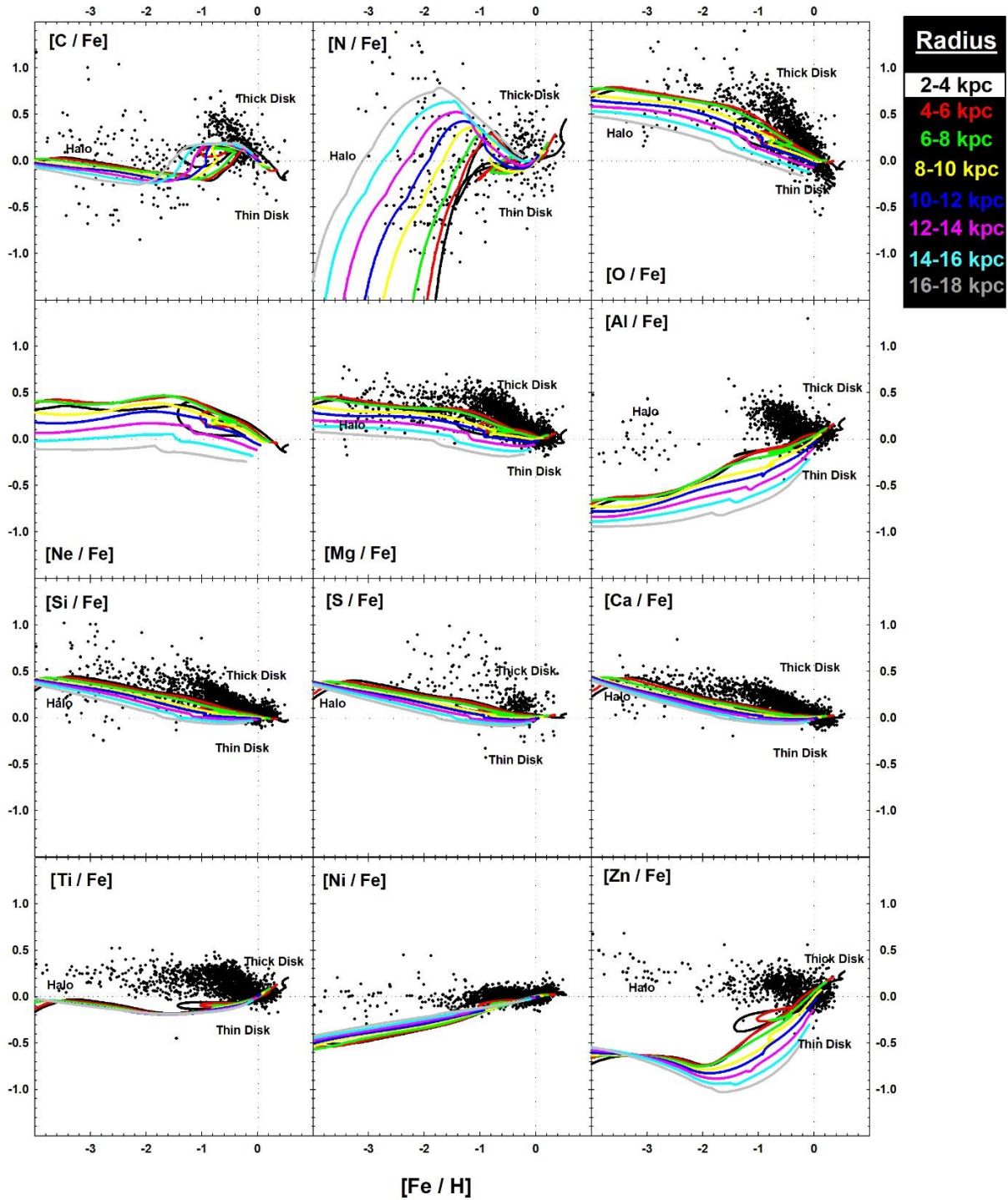

**Fig. 22.** Identical to Fig. 13 for **Model E**.

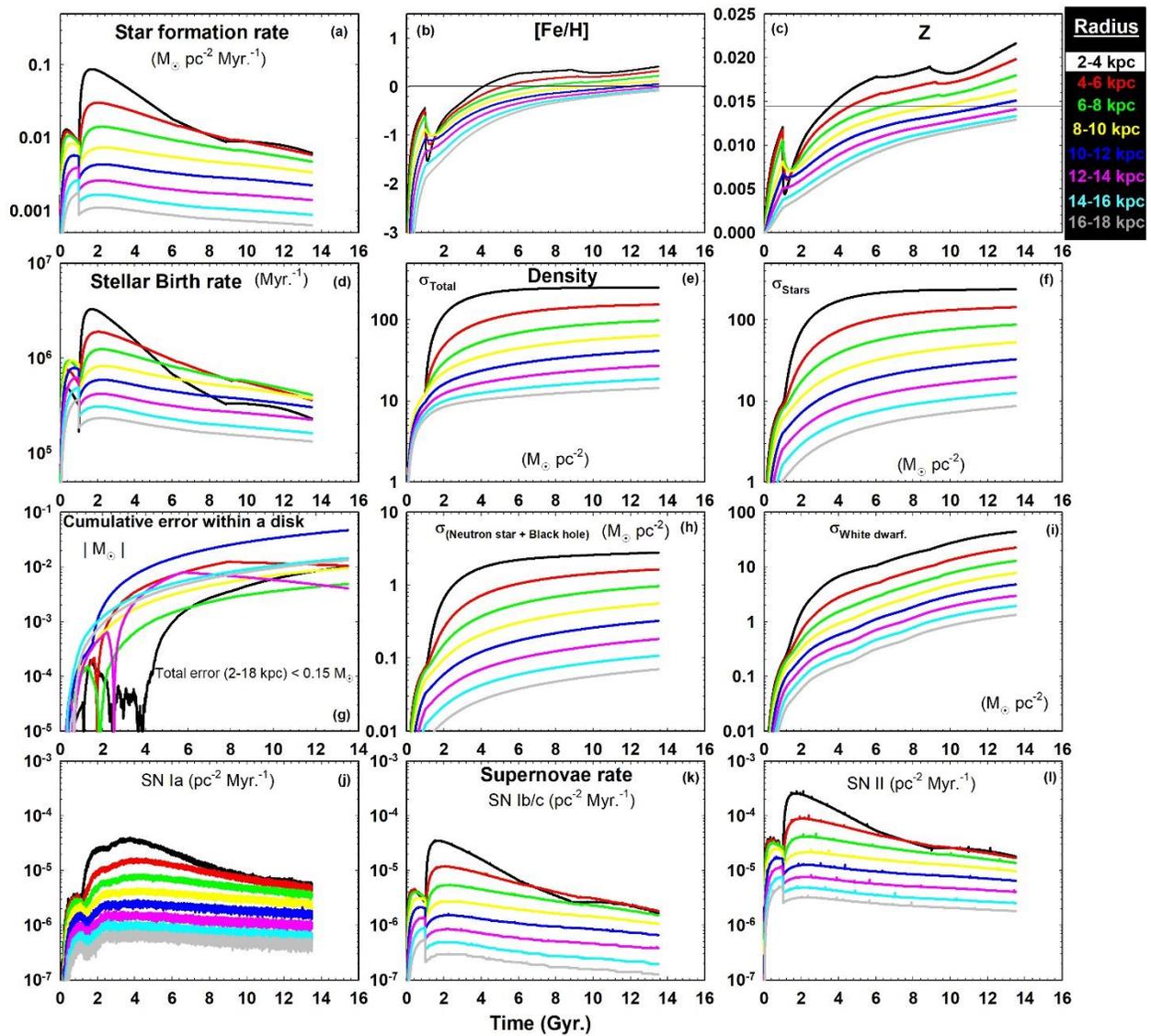

**Fig. 23.** Identical to Fig. 11 for *Model F*.

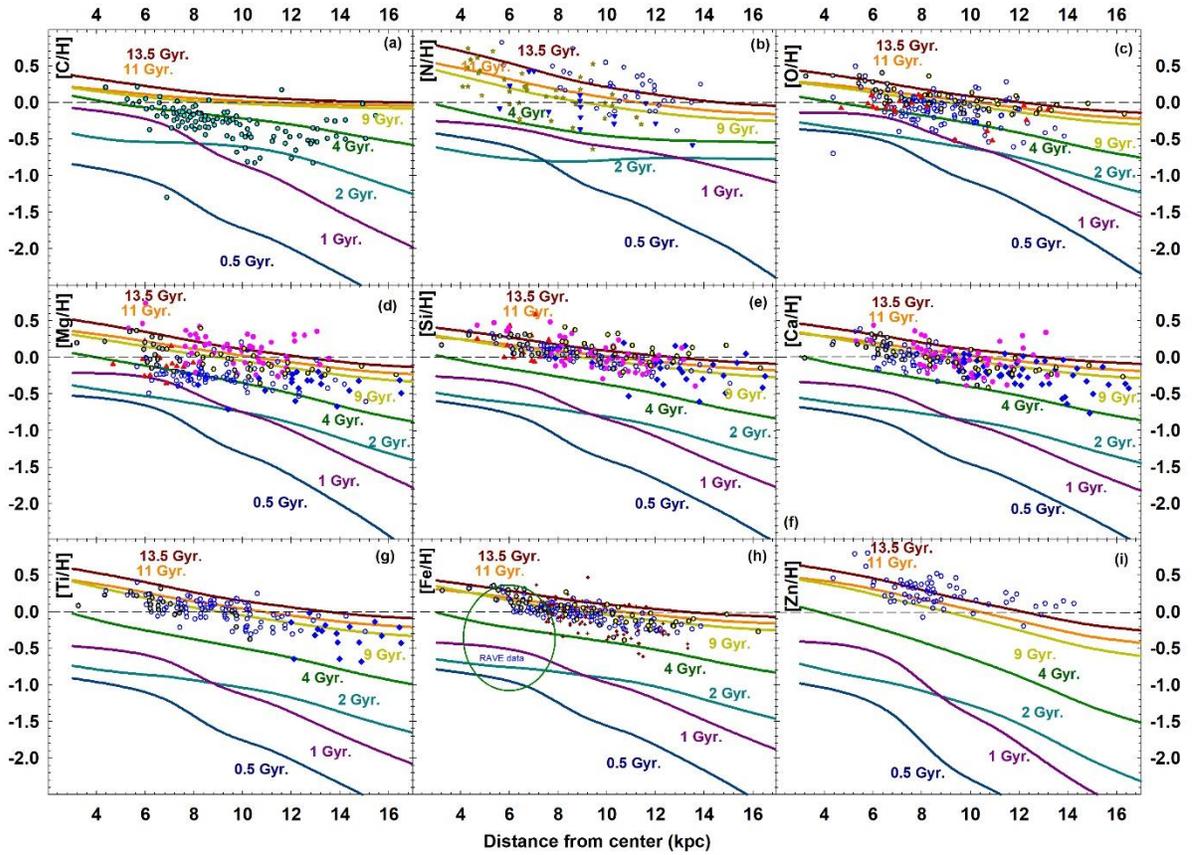

**Fig. 24.** Identical to Fig. 12 for *Model F*.

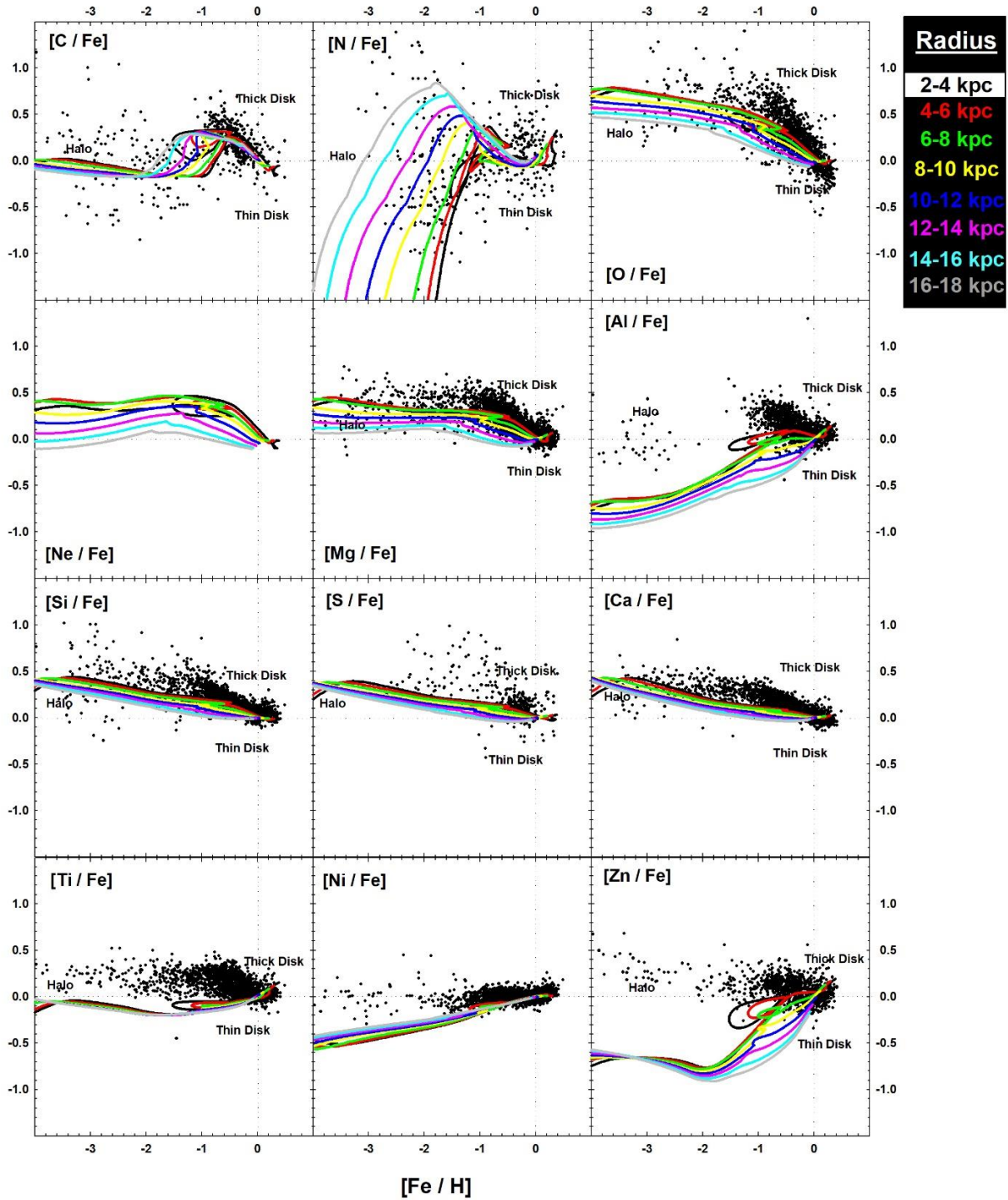

**Fig. 25.** Identical to Fig. 13 for *Model F*.

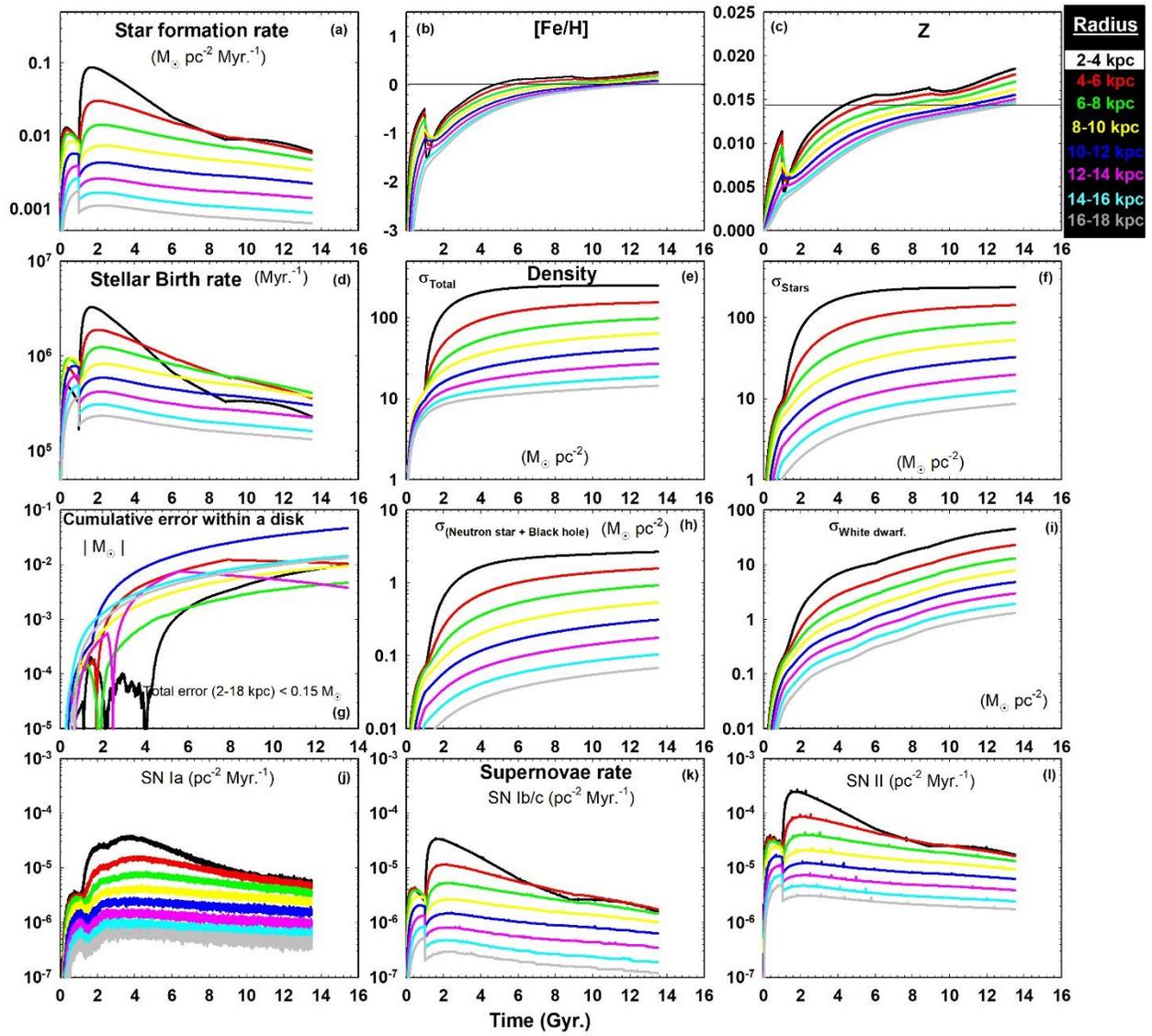

**Fig. 26.** Identical to Fig. 11 for *Model G*.

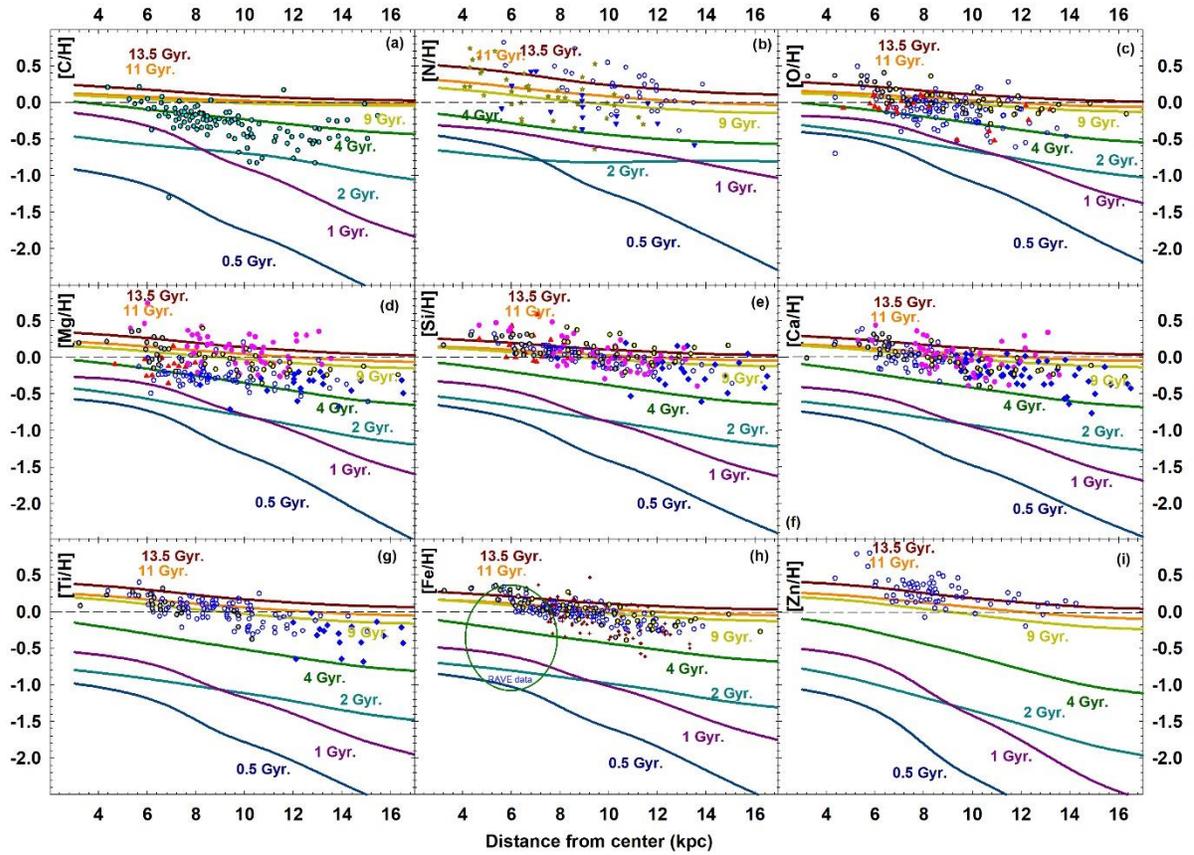

**Fig. 27.** Identical to Fig. 12 for ***Model G***.

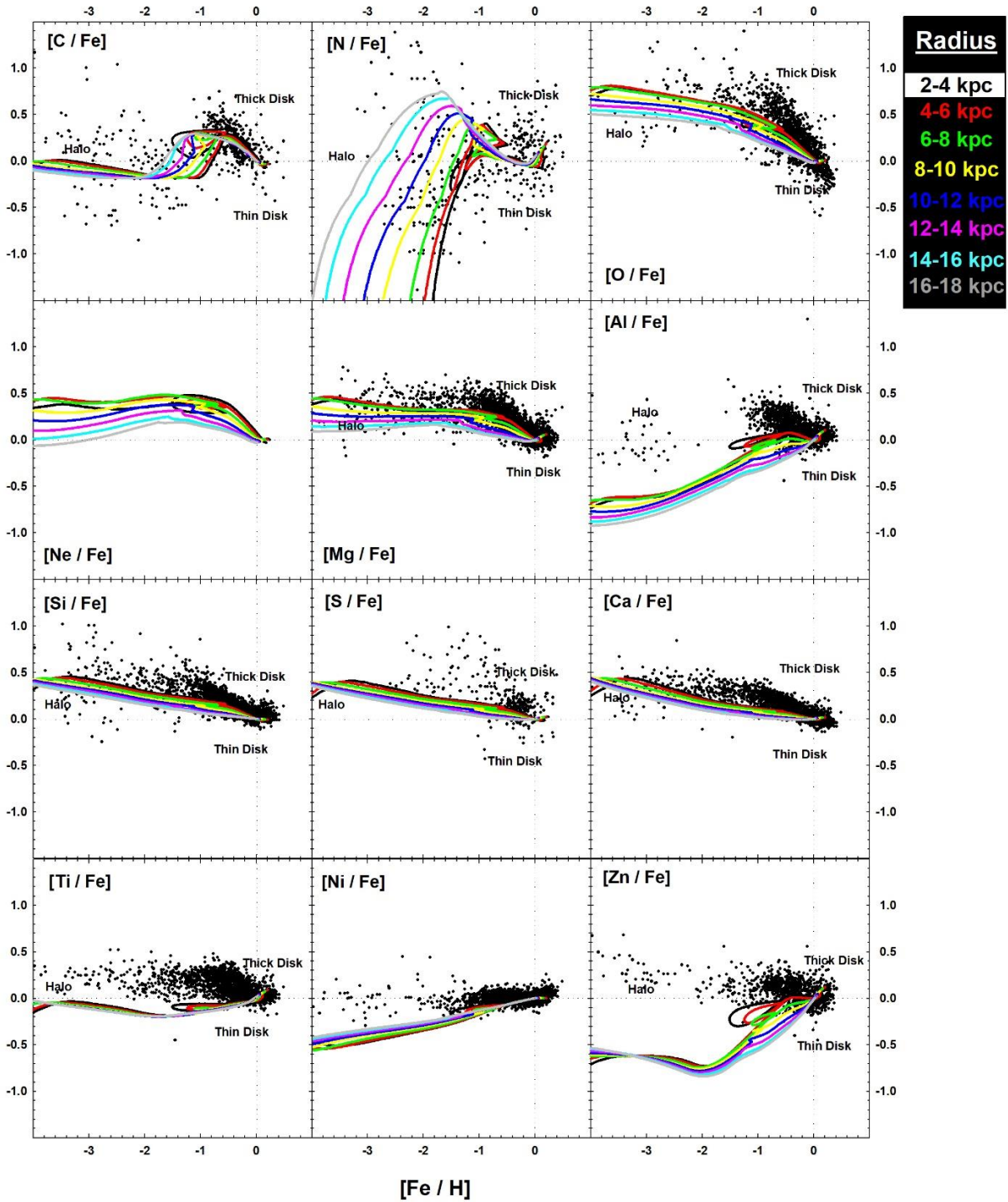

**Fig. 28.** Identical to Fig. 13 for *Model G*.

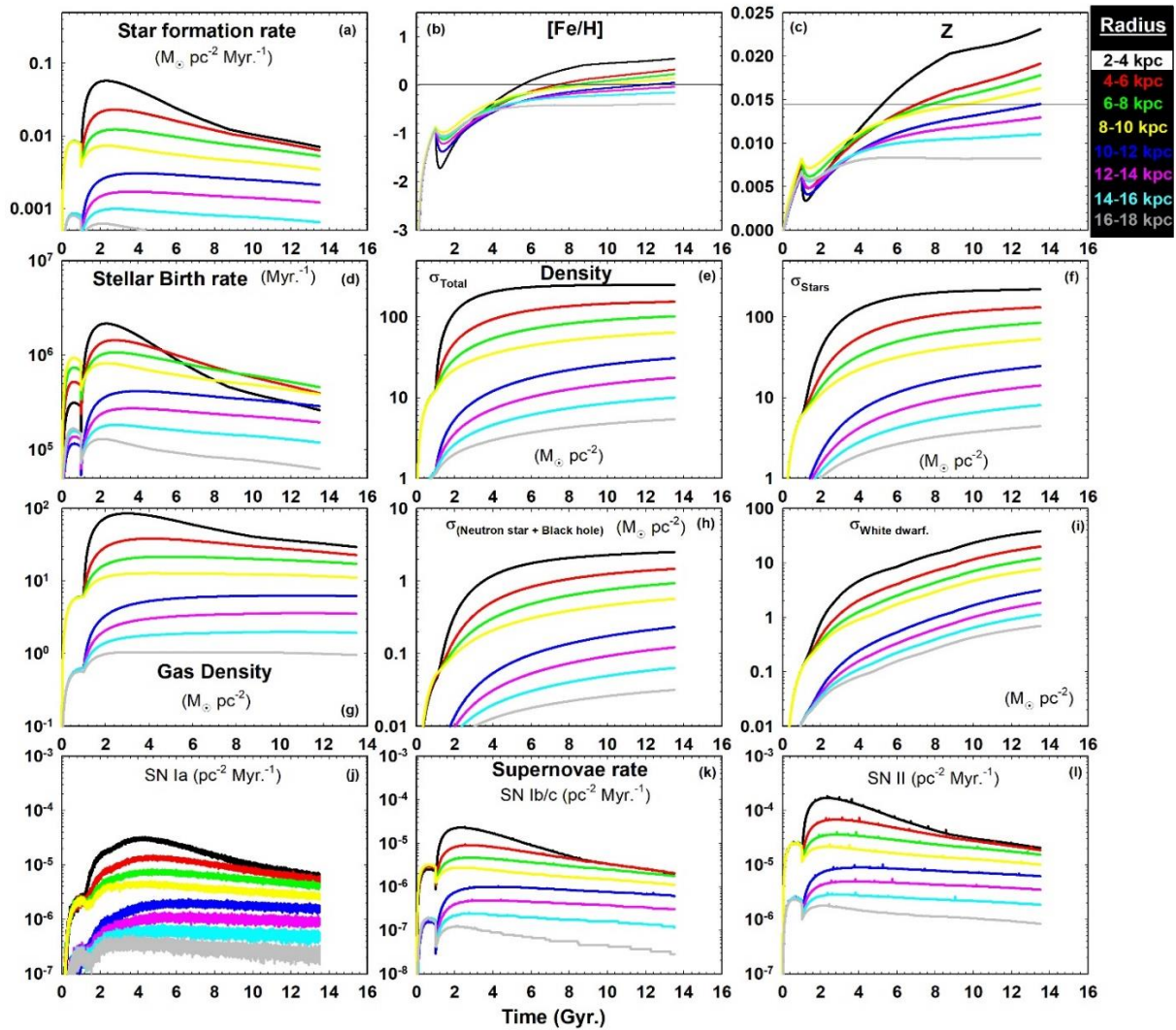

**Fig. 29.** The numerically deduced temporal evolution in case of ***Model I*** of, **a)** the star formation rate (M$_\odot$ pc$^{-2}$ Myr$^{-1}$), **b)** [Fe/H], **c)** the metallicity, '$Z$', **d)** the stellar birth rate (in Myr$^{-1}$), **e)** the total surface mass density (stars+interstellar gas), $\sigma_{Total}$ (M$_\odot$ pc$^{-1}$), **f)** the surface mass density of the various *live* stars $\sigma_{stars}$ (M$_\odot$ pc$^{-1}$), **g)** the gas surface mass density, $\sigma_{Gas}$ (M$_\odot$ pc$^{-1}$), **h)** the remnant surface mass density of Neutron stars + Black holes, **i)** the surface mass density of white dwarfs, are presented for the various annular rings. **j-l)** The deduced temporal evolution of the Supernovae (SN Ia SN II, and Ib/c) rates for the various annular rings are also presented.

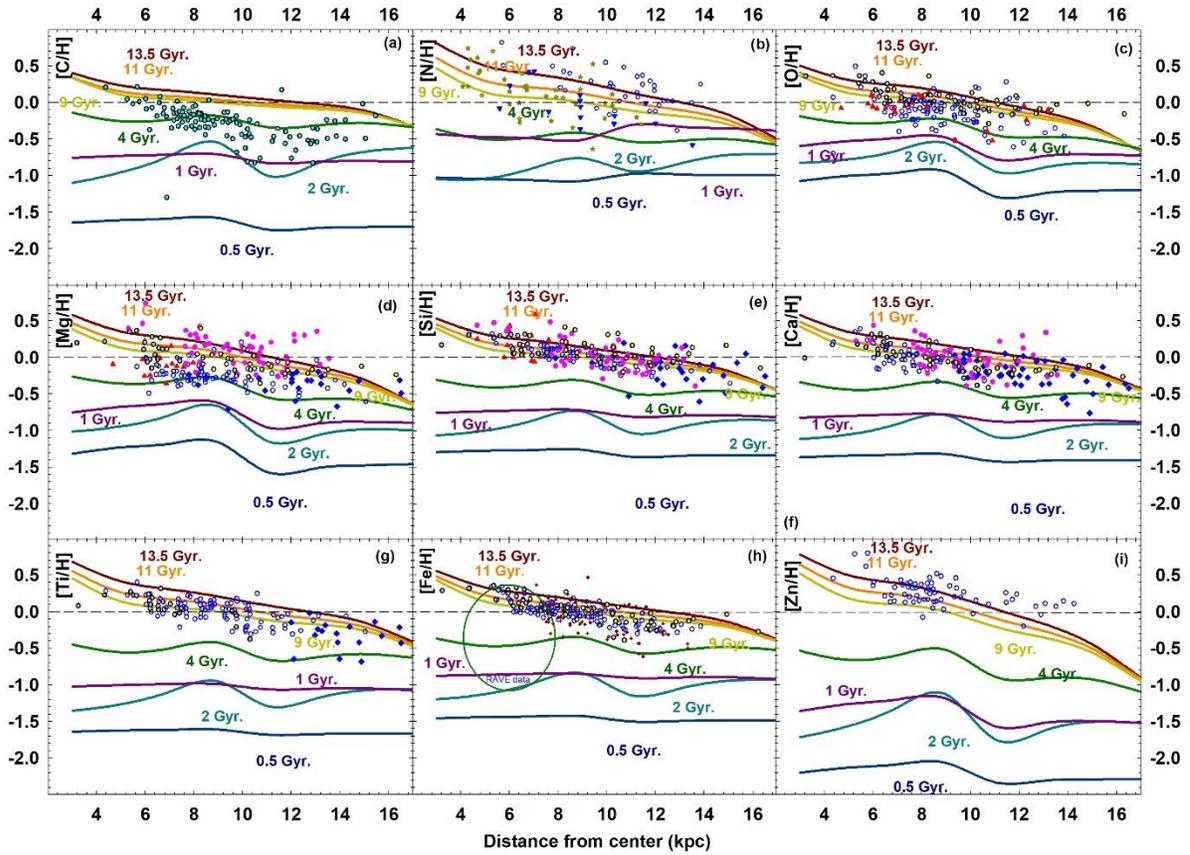

**Fig. 30.** Identical to Fig. 12 for *Model I*.

   Monthly Notices of the Royal Astronomical Society

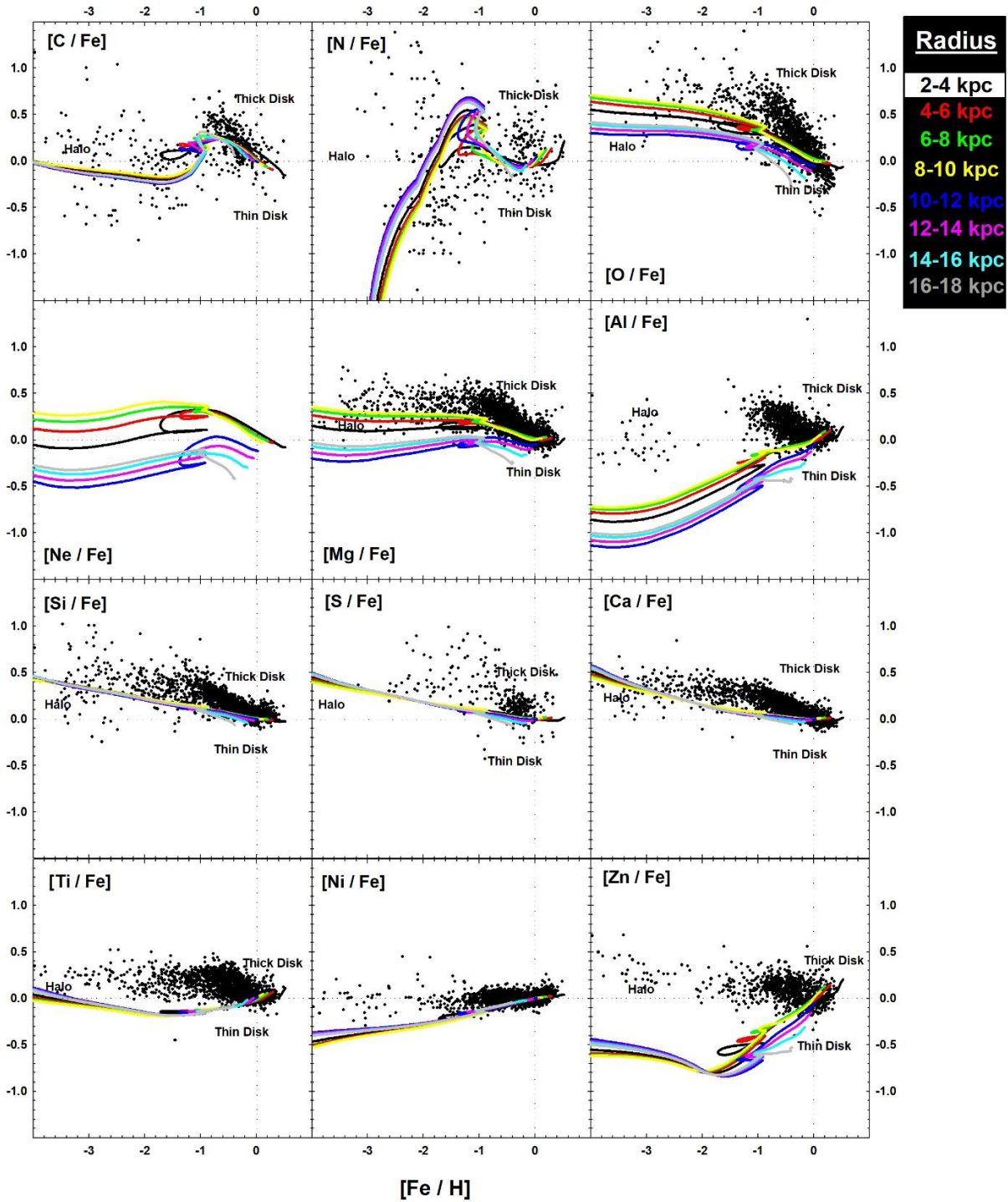

**Fig. 31.** Identical to Fig. 13 for *Model I*.